\documentclass[12pt]{article}
\usepackage{helvet,amsmath,amssymb,amsfonts,bm,yhmath,bbm} 
\newif\ifusesec
\usesectrue 
\usepackage{geometry}                
\usepackage{graphics}    
\usepackage{epstopdf}
\usepackage{helvet,amsmath,amssymb,amsfonts,amsthm,bm} 
\usepackage[latin1]{inputenc} 
\usepackage[T1]{fontenc} 
\usepackage[english]{babel}
\usepackage{graphicx,epsfig,stackrel} 
\usepackage[all]{xy}
\usepackage{hyperref}
\usepackage{tablefootnote}
\usepackage{accents}
\newcommand\sbullet[1][.5]{\mathbin{\vcenter{\hbox{\scalebox{#1}{$\bullet$}}}}}

\newcommand\dens[1]{\underaccent{ \sbullet}{#1}\mathstrut}
\newcommand{\ft}[2]{{\textstyle\frac{#1}{#2}}}
\numberwithin{equation}{section}
\begin{document}
\title{ Dynamical torsion  gravity backgrounds} 
\author{Ph. Spindel,\thanks{Electronic address : philippe.spindel@umons.ac.be}\\
Physique de l'Univers, Champs et Gravitation, UMONS,\\ 20 Place du parc, 7000 Mons, Belgium\\
Service de Physique th\'eorique, CP 225, ULB, \\Bld du Triomphe, 1050 Brussels, Belgium}
\maketitle

{\begin{flushright}{\it \`A la m\' emoire de  Jacques Hoeymans : \\ un ami, physicien et rationaliste \ldots\ \strut}\end{flushright}}

\begin{abstract}
 We write the field equations of torsion gravity theories and the Noether identity they obey directly in terms of metric and contorsion tensor components expressed with respect to natural coordinates, i.e. without using vierbien but Lagrange multipliers. Then we obtain explicit solutions of these equations, under specific ans\"atze for the contorsion field, by assuming the metric to be respectively of the Bertotti-Robinson, pp-wave, Friedmann-Lema\^itre-Robertson-Walker or static spherically symmetric type. Among these various solutions we obtain some of them have their contorsion tensor depending on arbitrary functions that did not influence their geometry. This raises questions about the predictability of the theory.
\end{abstract}
\section{Introduction}
Recently there has been renewed attention to modified gravity theories. The main motivations
for these endeavors are on the one hand purely theoretical (the quest of an unification of all forms
of interactions and matter and of a deeper understanding of the peculiarities of Einstein gravity
theory) and on the other hand dictated by the desire to offer alternative explanations for the recent cosmological
observations that have led to the introduction of hypothetical dark matter and energy. There are in fact several versions
of modified gravity \cite{baekler1986nonmetricity,blagojevic2013gauge,Clifton_2012,de_Rham_2014}. 
Of the many these theories, torsion gravity looks particularly interesting. It is a geometrical theory, based on a dynamical metric and a dynamical independent metric preserving connection, thus generalising general relativity. In their simplest expression they are obtained by adding to the metric and connection scalar curvature terms quadratic in the torsion and curvature tensor that for appropriate values of the coupling constants of these extra-terms provides physically sane models i.e. without ghost and tachyon  ( see Refs \cite{PhysRevD.21.3269,PhysRevD.24.1677,Hayashi1980gravity1,Hayashi1980gravity2,Hayashi1980gravity3,Hayashi1980gravity4} ). \\

Compared to the usual Einstein-Hilbert gravity, torsion gravity theories extend  the usual gravity
framework by modifying the infrared sector of the theory via the introduction, in addition to the massless
gauge spin 2 excitations, of positive and negative parity massive spin 2, 1 and zero modes. Such a theory was first
considered by Cartan \cite{cartan1923varietes, cartan1924varietes, cartan1925varietes}  (for a historical perspective of the matter and its developments
see for instance\cite{debever1979elie,hehl1976general}). Contrary to Palatini approach \cite{Palatini:1919aa} where the field variables are the
metric components and a torsionless connection, the variables used in
Einstein-Cartan-(Weyl\cite{weyl1950remark}-Sciama\cite{sciama1962analogy}-Kibble\cite{kibble1961lorentz}) are the coframe and spin-rotation coefficient components. In Cartan's works the Lagrangian was restricted
to the scalar curvature defined by the metric and an affine (but metric preserving) connection.
As a consequence the torsion degrees of freedom did not propagate but were determined by an
algebraic equation coupling them to the matter spin content. Accordingly the torsion variables
could be eliminated by the introduction of spin-spin matter interactions.

To obtain a propagating torsion, we have to introduce non-linear terms in the torsion tensor.
Of course this open the Pandora's box of ghosts and tachyonic modes. Remarkably, in
a set of seminal works, Sezgin and van Nieuwenhuizen \cite{PhysRevD.21.3269,PhysRevD.24.1677}and Hayashi and
Shirofuji \cite{Hayashi1980gravity1,Hayashi1980gravity2,Hayashi1980gravity3,Hayashi1980gravity4} have studied the most general invariant Lagrangian, at most quadratic in the torsion and curvature. They have analysed the perturbation spectrum around 
flat space and showed that there exists two classes (each ones depending on five parameters) of models without pathologies (i.e.
without ghosts and tachyons once expanded around the empty space configuration) that describe in addition to the usual massless spin 2 graviton,
massive spin 2, spin 1 and spin 0 excitations. The main difference between these two classes is the parity of their fluctuation field : $0^+$ and $2^-$ for the class $I$ models, and $0^-$ and $2^+$ for the class $I\! I$ models.

Usually these theories are formulated in terms of a coframe (vierbein) and the curvature tensor
built on a metric preserving affine connection. It is the latter that introduces the torsion. This
choice of variables is unavoidable when spinorial matter sources are taken into account. It offers
the advantage that the metric-preserving property of the connection is simply implemented by
the requirement that the connection 1-form is antisymmetric in its Lorentz indices, but requires
additional gauge fixing conditions to specify the  {\it a priori}   arbitrary 16 coframe components. It also
requires, for consistency at a classical level when fermionic fields are present, the use of Grassmannian variables which leads to a
triangular hierarchy of the field equations. Nevertheless at Grassmann degree zero the fermionic
variables did not play any r\^ole in the field equations. They may be ignored and in the absence of
bosonic matter the system is only driven by the geometrical variables.  { It is such configuration
that we will consider in this work : purely geometrical torsion gravity theories in a vacuum scheme (with a cosmological constant) obtained from  ghost- and tachyon-free Lagrangians \cite{PhysRevD.21.3269,PhysRevD.24.1677,Hayashi1980gravity1,Hayashi1980gravity2,Hayashi1980gravity3,Hayashi1980gravity4}. }

A glance on the Net, using the keyword {\it ``modified gravity"} in the title of papers provides around
a thousand of entries.    { During the years a huge number of exact solutions of modified gravity theories have been produced (see for instance Refs  \cite{adamowicz1980plane,baekler1988hamiltonian,blagojevic2013gauge,blagojevic2017generalized,obukhov2018poincare,Nikiforova_2018,PhysRevD.95.024013}). Also exact solutions have been considered, in the framework of the perfect fluid or the electromagnetic schemes; see for example \cite{zhytnikov1993conformally,minkevich2016towards}  and the references therein.  Moreover it is important to recall the so-called double duality method (see Refs \cite{Bakler:1980mu,baekler1983vacuum,mielke1984reduction,baekler1988hamiltonian,wallner1991exact, zhytnikov1996double}), which  reduces some dynamical equations to Bianchi identities and allows to obtain solutions analogous to the instantons of Yang-Mills theories. The purpose of this work is to provide some
new solutions, whose behaviour of some may be the indication of a weakness of (some sectors of) these theories : they are not predictable in the sense of the uniqueness of the Cauchy problem. We postpone for a future work a detailed  analysis of this crucial point. To build the  solutions we present we
formulate directly the field equations as Euler-Lagrange equations obtained from variation with
respect to the metric (instead of the vierbein) and the connection expressed in natural coordinates
(instead of spin coefficients).} 
Of course we will have to manage the metric preserving character
of the connection, but as we shall discuss later this can be done quite easily by introducing
Lagrange multipliers, whose constraint equations are trivially solved.

The paper is organised as follows. In section [\ref{Feqs}]  we establish the field equations using this last
approach and show their equivalence with those obtained from more common Cartan variables
(i.e. coframe and spin connection). We also briefly discuss the Noether identities satisfied by
the field equations, mainly because they constitute a useful check of their correctness and are a
key to establish, in some cases, the equivalence between the full set of equations of motion and
those obtained from a reduced Lagrangian obtained after the substitution of an ansatz in the
original one. In section [\ref{QuadLag}]  we particularise the general field equations to those resulting from a
Lagrangian quadratic in the torsion field components. Then we present some (new) solutions of
the field equations, obtained under different simplifying assumptions. 

The solutions we present hereafter offer miscellaneous interesting aspects. The first we discuss   are 
torsionless. It is known from a long time \cite{obukhov1989quadratic} that among the conformally flat geometry only de Sitter or anti-de Sitter spaces are solutions of the field equations, unless a special combination of the parameters of the models are related to the cosmological constant. We obtain a particular solution in the framework of class $I$ models under this constraint. Then we turn to torsionful solutions. To solve the field equations we make specific ans\"atze by restricting the expression of the metric (choosing a form that solves the field equation in absence of torsion) and by fixing the  {\it a priori}   non vanishing 
components of the contorsion tensor that are chosen in accordance with the metric symmetries.  By considering a Bertotti-Robinson geometry \cite{robinson1959solution,bertotti1959uniform}, which is a symmetric space \cite{cahen1968metriques}, we obtain solutions whose contorsion tensors involve arbitrary functions. Next we turn
to plane-fronted wave spacetimes \cite{Brinkmann} and deform them by adding contorsion. In the case of class $I$ solutions we obtain that the metric continues to define an Einstein space. For class $I\! I$ it is no more the case, illustrating the gravitational nature of the positive parity massive spin two. Next we turn to the simplest Friedmann-Lema\^itre-Robertson-Walker geometries and obtain various cosmological solutions. First we consider de Sitter solution in the framework of class $I$ models, under the same assumptions \cite{PhysRevD.95.024013} that have led to exclude time depending contorsion field configurations in the framework of 
class $I\! I$ models. Here again the difference between the two classes is illustrated. We obtain  a  time dependent contorsion configuration over a de Sitter geometry.  Then we turn to class $I\! I$ models. To go ahead  we freeze out the scalar modes and obtain more general solutions that those of de Sitter.  Some describe spaces evolving between an initial and a final singularity. But more interesting we also obtain a solution offering a metric everywhere regular, not de Sitter but interpolating between two de Sitter geometries. The metric of this solution is everywhere well defined  but nevertheless its domain of validity is restricted by a singularity in the contorsion field. To make an end we display, in the framework of class $I$ models a black hole configuration whose contorsion field depends on an arbitrary function, and show that such configuration may not appears in the framework of class $I\! I$ models.
In appendix [\ref{GBE}], we recall the construction of the Euler topological invariant (an higher dimensional generalisation of the Gauss-Bonnet invariant) and from it sketch a
proof of a quadratic identity (immediately extended to higher even dimensions), discovered by Bach \cite{Bach:1921aa} and Lanczos \cite{Lanczos1938}, that is satisfied by Riemann
curvature tensor and that we use during our work. Finally, for the readers convenience we summarise the conventions used by various authors that have inspired this work.

 \section{Torsion gravity field equations\label{Feqs}}
\subsection{Metric \& connection formalism}
The models are expressed in terms of two sets of variables. The metric components $g^{\mu\nu}$ and the affine connection components $A^\alpha_{{\, .\,} \beta\mu}$, defining a covariant derivative, denoted by ${\overline\nabla\!}$, that is assumed to be metric preserving :
\begin{align}{\overline\nabla\!}_\mu\,g^{\rho\sigma}:=\partial_\mu\,g^{\rho\sigma}+g^{\lambda\sigma}A^{\rho}_{{\, .\,} \lambda\mu}+g^{\rho\lambda }A^{\sigma}_{{\, .\,} \lambda\mu}=0\qquad .\label{metpres}
\end{align}
In what follows we will distinguished between the Levi-Civita  connection  (denoted as usual, with respect to a natural basis/coordinate system by $\Gamma^{\alpha}_{{\, .\,} \beta\mu}$), $\nabla$ denoting the covariant differential associated to it  and $R^{\alpha}_{\beta\mu\nu}=\partial_\mu\,\Gamma^{\alpha}_{{\, .\,} \beta \nu}+\dots $ its curvature tensor. The latter will be called Riemann-tensor in order to be distinguished from the curvature tensor obtained from the affine connection  whose components in a natural basis reads :
 \begin{align}&F^{\alpha}_{{\, .\,} \beta\mu\nu}=\partial_\mu\,A^{\alpha}_{{\, .\,} \beta  \nu}+A^{\alpha}_{{\, .\,} \rho \mu }A^{\rho}_{{\, .\,} \beta \nu }-\partial_\nu\,A^{\alpha}_{{\, .\,} \beta  \mu}+A^{\alpha}_{{\, .\,} \rho \nu }A^{\rho}_{{\, .\,} \beta \mu }\qquad ,\\
&\phantom{F^{\alpha}_{{\, .\,} \beta\mu\nu}}=R^{\alpha}_{{\, .\,} \beta\mu\nu} +{\nabla\!}_\mu K^\alpha_{{\, .\,} \beta\nu}-{\nabla\!}_\nu K^\alpha_{{\, .\,} \beta\mu}+ K^\alpha_{{\, .\,} \rho\mu}K^\rho_{{\, .\,} \beta\nu}-K^\alpha_{{\, .\,} \rho\nu}K^\rho_{{\, .\,} \beta\mu}\qquad .\label{F4R4}
\end{align}
 To be complete, we recall  the definitions   of the contorsion and torsion tensors components in natural coordinates. The first is obtained as the difference between the affine connection and the Levi-Civita connection~:
 \begin{align}K^\alpha_{{\, .\,}\beta\gamma}:=A^\alpha_{{\, .\,}\beta\gamma}-\Gamma^\alpha_{{\, .\,}\beta\gamma}\qquad .
 \end{align}
 The second is related to the antisymmetric part of the affine connection :
 \begin{align}T^\alpha_{{\, .\,} \beta\gamma}=A^\alpha_{{\, .\,}  \gamma\beta}-A^\alpha_{{\, .\,} \beta\gamma}=K^\alpha_{{\, .\,}  \gamma\beta}-K^\alpha_{{\, .\,} \beta\gamma}\qquad .\label{TK}\end{align}
 These two objects being obtained  from differences of connections are tensors. The flip of the indices in the torsion with respect to those of the connection is a reminiscence of the natural formalism to discuss these objects :  Cartan's exterior differential calculus. 
 
 The Lagrangian   we shall consider consists of two pieces. The usual Einstein-Hilbert Lagrangian  {\footnote{  {See Appendix [\ref{conv}] for some specific conventions used in this work.}} }(including a bare cosmological constant)  with a coupling constant $c_R$~: 
 \begin{align}
 \dens{ \mathcal L}_{E-H}:=\sqrt{-g}(c_R\,R-2\,\Lambda) \qquad ,\label{LagEH}
 \end{align}
 and the ``connection matter Lagrangian'' : 
  \begin{align}
 &\dens {\mathcal L}_F(g^{\mu\nu},F^\alpha_{{\, .\,}\beta\gamma\delta})=\sqrt{-g}\,L_F(g^{\mu\nu},F^\alpha_{{\, .\,}\beta\gamma\delta})\label{LagPgF}
 \end{align}
  that we assume only to depend on the metric (but not on its derivative) and on the curvature tensor. The introduction of a cosmological constant is disputable. One primary aim of torsion gravity models is to provide a dynamical origin of the acceleration of the Universe. Moreover we also have to remind the reader that some conditions leading to the absence of tachyons or ghosts have to be reconsidered on curved backgrounds, in particular on an anti-deSitter background \cite{breitenlohner1982positive,breitenlohner1982stability}. Nevertheless the consistency of the models around these backgrounds has been established in Ref. ~\cite{PhysRevD.80.104031} and extended in Ref.  \cite{Nikiforova2009infrared} to weakly curved torsionless Einstein backgrounds.

 Thus the total Lagrangian is given by the sum $ \dens{\mathcal L}_{E-H}+ \dens{\mathcal L}_F$. The field equations are obtained by varying this Lagrangian with respect to $g^{\mu\nu}$ and $A^{\alpha}_{{\, .\,} \beta  \mu}$, taking into account the metric preserving assumption Eq. (\ref{metpres}).
 This condition is implemented with the help of Lagrange multipliers~: $\dens\lambda^\alpha_{{\, .\,} \beta\gamma}=\dens\lambda^\alpha_{{\, .\,} (\beta\gamma)}$. Thus the complete Lagrangian, that depends on the metric and connection components and their derivatives, read as  :
\begin{align}
 &\dens {\mathcal L}_P =\sqrt{-g}(L_{E-H}+L_F)+\dens\lambda^{\alpha}_{{\, .\,} \beta\gamma}{\overline\nabla\!}_\alpha g^{\beta\gamma}\qquad .
\end{align}
Let us define auxiliary quantities :
\begin{align}& {\Delta}\mathstrut _{\mu\nu}:=-\ft12 g_{\mu\nu} L _{F} +\left . \frac{\partial L_F}{\partial g^{\mu\nu}}\right\vert_{F^{\alpha}_{{\, .\,} \beta\gamma\delta}}= {\Delta}\mathstrut _{(\mu\nu)}\qquad ,\label{Dmn}\\
& Z_\alpha^{{\, .\,} \beta\gamma\delta}:=\left .\frac{\partial  L_F}{\partial F^{\alpha}_{{\, .\,} \beta\gamma\delta}}\right \vert_{g^{\mu\nu}}=   Z_\alpha^{{\, .\,} \beta[\gamma\delta]}\qquad ,\\
&\Delta_\alpha^{{\, .\,} \beta\gamma}=\ft 1{2\,\sqrt{-g}}\frac {\delta \dens L_F}{\delta A^\alpha_{{\, .\,} \beta\gamma}}=({\overline\nabla\!}_\delta Z_\alpha^{{\, .\,} \beta\gamma\delta}+\ft 12 Z_\alpha^{{\, .\,} \beta\rho\sigma}\,T_{{\, .\,} \rho\sigma}^\gamma-Z_\alpha^{{\, .\,} \beta\gamma\rho}\,T_{  \rho})\\
&\phantom{\Delta_\alpha^{{\, .\,} \beta\gamma}=\ft 1{2\,\sqrt{-g}}\frac {\delta \dens L_F}{\delta A^\alpha_{{\, .\,} \beta\gamma}}}={\nabla\!}_\delta Z_\alpha^{{\, .\,} \beta\gamma\delta}-K^\rho_{{\, .\,} \alpha\delta}Z_\rho^{{\, .\,} \beta\gamma\delta}+K^\beta_{{\, .\,} \rho\delta}Z_\alpha^{{\, .\,} \rho\gamma\delta}\qquad .
\end{align}
The variational derivative   (taking into account the Lagrangian multipliers) are :
\begin{align} 
&\frac{\delta\dens {\mathcal L}_P}{\delta g^{\mu\nu}}=+\dens{\Delta}_{\mu\nu}-\ft 12 g_{\mu\nu}\,\dens\lambda^{\alpha}_{{\, .\,} \beta\gamma}{\overline\nabla\!}_\alpha g^{\beta\gamma}-{\overline\nabla\!}_\alpha \dens{\lambda}\mathstrut^\alpha_{{\, .\,} \mu\nu}+T_\alpha\,\dens \lambda\mathstrut^\alpha_{{\, .\,} \mu\nu}\qquad , \\
&\frac{\delta\dens {\mathcal L}_P}{\delta A^{\alpha}_{{\, .\,} \beta\gamma}}=2\,\dens\Delta\mathstrut^{{\, .\,} \beta\gamma}_{\alpha}+\dens\lambda\mathstrut^{\gamma{\, .\,} \beta}_{{\, .\,} \alpha}+ \dens\lambda\mathstrut\mathstrut^{\gamma\beta}_{{\, .\,} {\, .\,} \alpha}\label{coneqs} \qquad ,
\end{align}
The  Lagrange multiplier Euler equations imply the antisymmetry of the contorsion tensor $K^{\alpha}_{{\, .\,}\beta\mu}$~:
\begin{align}A^{\alpha}_{{\, .\,}\beta\mu}=\Gamma^{\alpha}_{{\, .\,}\beta\mu}+K^{\alpha}_{{\, .\,}\beta\mu}\qquad\text{with}\qquad K_{\alpha\beta\mu}:=g_{\alpha\gamma}K^{\gamma}_{{\, .\,}\beta\mu}=K_{[\alpha\beta]\mu}
\end{align}
while the connection Euler equation fix the Lagrange multiplier expression $\lambda\mathstrut^\alpha_{{\, .\,} \mu\nu}$~ :
\begin{align}& \lambda^{\gamma\alpha\beta }=\Delta^{(\alpha\beta)\gamma}\qquad .
\end{align}
The remaining field equations are :
\begin{align} &\dens {{\mathcal S}}_{\,\gamma}^{{\, .\,} \alpha\beta}\equiv (\dens{\Delta}\mathstrut_\gamma^{{\, .\,}\alpha\beta}-\dens{\Delta}\mathstrut_{{\, .\,}\gamma{\, .\,}}^{\alpha{\, .\,}\beta})=0 \qquad ,\label{Kmunu}\\
&(\text{i.e. }\quad
\Delta^{[\alpha\beta]\gamma}={\nabla\!}_\delta Z^{[\alpha \beta]\gamma\delta}-K^\alpha_{{\, .\,} \rho\delta}Z^{[\beta\rho]\gamma\delta}+K^\beta_{{\, .\,} \rho\delta}Z^{[\alpha\rho]\gamma\delta}=0)\qquad ,\\
&\dens {\mathcal E}_{\mu\nu}\equiv \sqrt{-g}{\big(}c_R(R_{\mu\nu} -\ft 12\,g_{\mu\nu}\,R)+\Lambda\,g_{\mu\nu}\big)-\dens {\mathcal T}_{\mu\nu}=0\qquad .\label{Emunu}
\end{align}
with :
\begin{align}&\dens {\mathcal T}_{\mu\nu}=-\Big(\dens{\Delta}_{\mu\nu}+ ({\overline\nabla\!}_\alpha \dens{\Delta}^{{\, .\,}{\, .\,}\alpha}_{(\mu\nu)} -T_\alpha \dens\Delta^{{\, .\,}{\, .\,}\alpha}_{(\mu\nu)})\Big)\qquad .
\end{align}
Let us notice that on-shell :
\begin{align}{\overline\nabla\!}_\rho {\Delta}^{  \alpha\beta\rho}-T_\rho \,{\Delta}^{  \alpha\beta\rho}=\ft 12 \big (Z^{\alpha\rho\mu\nu}F^{\beta}_{{\, .\,} \rho\mu\nu}+Z^{\rho\beta\mu\nu}F^{\alpha}_{{\, .\,} \rho\mu\nu}\big) \label{NabDelZF}\qquad .
\end{align}
Accordingly, defining the symmetric part  
\begin{align}S^{\alpha\beta\mu\nu}:=Z^{(\alpha\beta)\mu\nu}\qquad ,
\end{align}
we obtain the symmetric contorsion energy-momentum tensor :
\begin{align}\mathcal{T}_{\alpha\beta}=-(\Delta_{\alpha\beta}+F_{(\alpha}^{\,{\, .\,}\mu\nu\rho}\,S_{\beta)\mu\nu\rho})  \qquad .\label{STmunu}
\end{align}
This last expression shows that the Einstein equations involve only polynomials of the curvature, but no derivatives of it ( contrary to the connection equations that involve first derivatives of the curvature tensor). Thus in general the field equations will involves at most third order derivatives of the metric components and second order derivatives of the contorsion components.\\
To make an end to this section let us mention that for a Lagrangian having the structure given by Eqs. [\ref{LagEH}, \ref{LagPgF}] the same field equations (Eqs [\ref{Kmunu}, \ref{Emunu}]) are obtained if the metric and the contorsion are taken as independent field variables.
 \subsection{ Noether identities ( A reminder )}
Let us briefly recall the essence of Noether identity applied to an invariant Lagrangian density $\dens {\mathcal L}$. There are two relevant such identities.  We restrict ourselves to Lagrangians like those here considered, {i.e. } such that $\dens {\mathcal L}$ depends at most on the second derivative of   fields $Q_\omega$ whose variations are tensors and whose Lie derivatives involve at most  the second derivative of the generator $\xi^\alpha$ of the infinitesimal coordinate change : 
 \begin{align}&{\frak{L}}_\xi Q_\omega=:\xi^\lambda\partial_\lambda Q_\omega + c^{\lambda_1}_{\omega\vert\mu}\partial_{\lambda_1}\xi^\mu+c^{(\lambda_1\lambda_2)}_{\omega\vert\mu}\partial^2_{\lambda_1\lambda_2}\xi^\mu\qquad .
 \end{align}
Let us define~:
 \begin{align} 
&P^\omega:=\frac{\partial \dens L}{\partial Q_\omega} \qquad,\qquad P^{\omega\vert \alpha}:=\frac{\partial \dens L}{\partial Q_{\omega,\alpha}}\qquad,\qquad P^{\omega\vert \alpha\beta}:=\frac{\partial \dens L}{\partial Q_{\omega,\alpha\beta}}\qquad .
\end{align}

The fundamental Noether identity \cite{Noether1918, kosmann20n6oether} reads :
  {\begin{align}\frac{\delta \dens L}{\delta Q_\omega}{\frak{L}}_\xi Q_\omega+\partial_\alpha\big((P^{\omega\vert \alpha}-\partial_\beta P^{\omega\vert\alpha\beta}){\frak{L}}_\xi Q_\omega+P^{\omega\vert \alpha\beta}\partial_\beta ({\frak{L}}_\xi Q_\omega)-\xi^\alpha\,\dens L\big)\equiv 0\qquad .
\end{align}}
Expanding it with respect to the arbitrary field $\xi^\mu$ and its derivatives we obtain from the invariance of the Lagrangian :
\begin{align*}&\frac{\delta \dens L}{\delta Q_\omega}\partial_\mu Q_\omega +\partial_\alpha(P^{\omega\vert\alpha}\partial_\mu   Q_\omega-\delta_\mu^\alpha\,\dens L)+P^{\omega\vert\alpha\beta}\partial^3_{\alpha\beta\mu}Q_\omega-\partial^2_{\alpha\beta}P^{\omega\vert\alpha\beta}\partial _{ \mu}Q_\omega\equiv 0\\
&\frac{\delta \dens L}{\delta Q_\omega}\,c^{\lambda_1}_{\omega\vert \mu}+\partial_\alpha(P^{\omega\vert\alpha}\, c^{\lambda_1}_{\omega\vert\mu})+P^{\omega\vert{\lambda_1}}\, \partial_\mu Q_{\omega}-\delta^{\lambda_1}_\mu\,\dens L+2\,P^{\omega\vert\alpha\lambda_1}\partial^2_{\alpha\mu}Q_\omega+P^{\omega\vert\alpha\beta}\partial^2_{\alpha\beta}c_{\omega\vert\mu}^{\lambda_1}-\partial^2_{\alpha\beta}P^{\omega\vert\alpha\beta}\,c_{\omega\vert\mu}^{\lambda_1}\equiv 0\\
&\frac{\delta \dens L}{\delta Q_\omega}\,c^{\lambda_1\lambda_2}_{\omega\vert \mu}+\partial_\alpha(P^{\omega\vert\alpha}\, c^{\lambda_1\lambda_2}_{\omega\vert\mu})+P^{\omega\vert(\lambda_1}\, c^{\lambda_2)}_{\omega\vert\mu} +2\,P^{\omega\vert \alpha(\lambda_1 }\partial_\alpha c_{\omega\vert \mu}^{\lambda_2)}+P^{\omega\vert \lambda_1 \lambda_2}\partial_\mu Q_{\omega }-\partial^2_{\alpha\beta}P^{\omega\vert\alpha\beta}\,c_{\omega\vert\mu}^{\lambda_1\lambda_2}\\
&+P^{\omega\vert\alpha\beta}\partial^2_{\alpha\beta}c_{\omega\vert\mu}^{\lambda_1\lambda_2}\equiv 0\\
&2\,P^{\omega\vert\alpha(\lambda_1}\partial_\alpha c_{\omega\vert \mu}^{\lambda_2\lambda_3)}+P^{\omega\vert(\lambda_1\lambda_2}c_{\omega\vert \mu}^{ \lambda_3)}+P^{\omega\vert(\lambda_1 }c_{\omega\vert \mu}^{\lambda_2\lambda_3)}\equiv 0\\
&P^{\omega\vert(\lambda_1\lambda_2}c_{\omega\vert \mu}^{\lambda_2\lambda_4)}\equiv 0
\end{align*}
Let us notice that the first identity says that $\dens L$ cannot depends explicitly on the coordinates. The second exhibits the link between the symmetric and the canonical energy-momentum tensors. The next ones express symmetry properties.\\
By combining all these identities we obtain N\oe ther's famous second theorem~:
\begin{align}&\frac{\delta \dens L}{\delta Q_\omega}\partial_\mu Q_\omega - \partial_{\lambda_1}{\big(}\frac{\delta \dens L}{\delta Q_\omega}\,c^{\lambda_1}_{\omega\vert \mu}\Big)+\partial^2_{\lambda_1\lambda_2}\Big(\frac{\delta \dens L}{\delta Q_\omega}\,c^{(\lambda_1\lambda_2)}_{\omega\vert \mu}\Big)\equiv 0\qquad .\label {Noe2th}
\end{align}

To apply Noether theorem in the framework of this work we have to make use of the expression of the Lie derivative of the metric~:
\begin{align}{\frak{L}}_\xi g^{\alpha\beta}=\xi^\mu\partial_\mu g^{\alpha\beta}-g^{\mu\beta}\partial_\mu \xi^\alpha-g^{\alpha\mu }\partial_\mu \xi^\beta=-({\nabla\!}^\alpha\xi^\beta+{\nabla\!}^\beta\xi^\alpha)\qquad .
\end{align}
and of  the connection \cite{lichnerowicz1955theories} :
\begin{align}{\frak{L}}_\xi A^\alpha_{\beta\gamma}=\xi^\mu\partial_\mu A^\alpha_{\beta\gamma}+ A^\alpha_{\beta\mu}\partial_\gamma \xi^\mu- A^\mu_{\beta\gamma}\partial_\mu \xi^\alpha+ A^\alpha_{\mu\gamma}\partial_\beta \xi^\mu+\partial^2_{\beta\gamma}\xi^\alpha\qquad .\label{LieDA1}
\end{align}
Note that this Lie derivative (\ref{LieDA1}) also defines a tensor since the difference of two connections is a tensor. Indeed it can be written as~:
\begin{align}&{\frak{L}}_\xi A^\alpha_{\beta\gamma}= {\frak{L}}_\xi \Gamma^\alpha_{\beta\gamma}+{\frak{L}}_\xi K^\alpha_{\beta\gamma}\qquad,\qquad{\frak{L}}_\xi \Gamma^\alpha_{\beta\gamma}={\nabla\!}_\beta{\nabla\!}_\gamma \xi^\alpha+R^{\alpha}_{{\, .\,} \gamma\sigma\beta}\xi^\sigma
\end{align}
 or in a more cumbersome expression (that we shall not display) using the ${\overline\nabla\!}$ operator and the torsion tensor.

In the context of this work  we obtain  from Eq. [\ref{Noe2th}]~: 
\begin{align}\dens {\mathcal T}_{\alpha\beta}\partial_\mu g^{\alpha\beta}+2\partial_\lambda\dens {\mathcal T}^\lambda_{\mu}\equiv\dens {{\mathcal S}}_\gamma^{{\, .\,} \alpha\beta}\partial_\mu A^{\gamma}_{{\, .\,} \alpha\beta}-\partial_\lambda {\big(}\dens {{\mathcal S}}_\gamma^{{\, .\,} \alpha\beta}( A^{\gamma}_{{\, .\,} \alpha\mu}\,\delta^\lambda_\beta+ A^{\gamma}_{{\, .\,} \mu\beta}\,\delta^\lambda_\alpha- A^{\lambda}_{{\, .\,} \alpha\beta}\,\delta^\gamma_\mu)\big)+\partial^2_{\alpha\beta}\dens {{\mathcal S}}_\mu^{{\, .\,} \alpha\beta}
\end{align}
that can be rewritten, using the Levi-Civita connection, in an explicitly covariant form :
\begin{align}&{\nabla\!}_\mu {\mathcal T}^\mu_\alpha\equiv \frac 12 \Big({\nabla\!}_\mu{\nabla\!}_\nu {{\mathcal S}}^{{\, .\,} \mu\nu}_\alpha -{{\mathcal S}}^{{\, .\,} \mu\nu}_\rho\,R^{\rho}_{{\, .\,} \mu\nu\alpha}-{\nabla\!}_\mu({{\mathcal S}}_\sigma^{{\, .\,} \mu\nu}\,K^{\sigma}_{{\, .\,} \alpha\nu})-{\nabla\!}_\nu({{\mathcal S}}_\sigma^{{\, .\,} \mu\nu}\,K^{\sigma}_{{\, .\,} \mu\alpha})+{\nabla\!}_\sigma({{\mathcal S}}_\alpha^{{\, .\,} \mu\nu}\,K^{\sigma}_{{\, .\,} \mu\nu})\nonumber\\
&\phantom{{\nabla\!}_\mu {\mathcal T}^\mu_\alpha\equiv \frac 12 \Big(}+{{\mathcal S}}_\rho^{{\, .\,} \mu\nu}{\nabla\!}_\alpha K^{\rho}_{{\, .\,} \mu\nu}\Big)\quad .\label{BianchiId}
\end{align}
Accordingly, as expected, on any background, the connection energy momentum tensor ${\mathcal T}^\beta_\alpha$ becomes divergenceless when the connection field equations~: ${{\mathcal S}}_\alpha^{{\, .\,} \beta\gamma}=0$ are satisfied.
\subsection{Coframe \& spin-connection formalism}
 Usually, authors prefer to use Cartan formalism to discuss torsion  gravity models. Their starting point is a coframe $\{\underline{e}^{\hat a}=e^{\hat a}_\mu\,dx^\mu\}$ defining the metric as~:
  {\begin{align}g_{\mu\nu}=\eta_{\hat a\hat b}e^{\hat a}_\mu\, e^{\hat b}_\nu\qquad ,
\end{align}}
where $\eta_{\hat a\hat b}$ are  (constant) components of a Minkowskian metric and $A_{{\hat a}{\hat b}\mu}$ 
the coframe components of a metrical connection~: 
\begin{align}A_{{\hat a}{\hat b}\mu}=A_{[{\hat a}{\hat b}]\mu}\qquad .
\end{align} 
Obviously the main advantage of this approach rests in this relation which encodes algebraically the metrical consistency of the connection.\\
The Einstein-Hilbert Lagrangian depends on the coframe components and their first and second derivative. The matter Lagrangian density depends on the coframe and connection components and the first derivative of the latter~:
\begin{align}\dens {\mathcal L}_C=e\,L(\eta^{\hat a\hat b}e_{\hat a}^\mu e_{\hat b}^\nu, e_{\hat a}^\alpha\,e^{\hat b}_\beta \,F^{\hat a}_{{\, .\,} \hat b\mu\nu})\qquad,\qquad (e:=\det [e^{\hat a}_\mu])
\end{align}
 via the curvature tensor components~:
\begin{align}&F^{\hat a}_{{\, .\,} {\hat b}\mu\nu}=\partial_\mu A ^{\hat a}_{{\, .\,}{\hat b}\nu}-\partial_\nu A^{\hat a}_{{\, .\,} {\hat b}\nu}+A^{\hat a}_{{\, .\,} {\hat c}\mu}\,A^{\hat c}_{{\, .\,} {\hat b}\nu}-A^{\hat a}_{{\, .\,} {\hat c}\nu}\,A^{\hat c}_{{\, .\,} {\hat b}\mu}\qquad .
\end{align}
 
 The variational derivatives with respect to them read :
\begin{align}&\frac{\delta\dens{\mathcal L}_C}{\delta e^{\hat a}_\mu}=e_{\hat a}^\alpha (-2\,\dens \Delta_\alpha^\mu +Z_\sigma^{{\, .\,} \mu\gamma\delta}F^\sigma_{{\, .\,} \alpha \gamma\delta}-Z_\alpha^{{\, .\,} \sigma\gamma\delta}F^\mu_{{\, .\,} \sigma \gamma\delta})=\dens \Theta_{\hat a}^\mu\\
&\frac
{\delta\dens{\mathcal L}_C}{\delta A^{\hat a}_{{\, .\,} \hat b\mu}}=e_{\hat a}^\alpha\,e^{\hat b}_\beta  (\dens{\Delta}\mathstrut_\alpha^{{\, .\,} \beta\mu}-\dens{\Delta}\mathstrut_{{\, .\,}\alpha {\, .\,}}^{\beta{\, .\,}\mu})
=e_{\hat a}^\alpha\,e^{\hat b}_\beta \,\dens {{\mathcal S}}_{\alpha}^{{\, .\,} \beta \mu}
\end{align}
Accordingly, by denoting~:
\begin{align}\dens \Theta^{\nu\mu}:=e^{\hat a\nu}\dens \Theta_{\hat a}^\mu
\end{align}
we remark that (see Eqs [\ref{STmunu}, \ref{NabDelZF}])~:
\begin{align}&\dens \Theta^{(\nu\mu)}=\dens {\mathcal T}_{}^{(\nu\mu)}\\
&\dens \Theta^{[\nu\mu]}=(\dens Z^{[\sigma\nu]\gamma\delta}F^{\mu}_{{\, .\,} \sigma\gamma\delta}-\dens Z^{[\sigma\mu]\gamma\delta}F^{\nu}_{{\, .\,} \sigma\gamma\delta})\\
&\phantom{\dens {\mathcal E}^{[\nu\mu]}}=-{\overline\nabla\!}_\gamma \dens {{\mathcal S}}^{[\nu\mu]\gamma}+T_\gamma\,\dens {{\mathcal S}}^{[\nu\mu]\gamma}
\end{align}
which explicits the equivalence of the metric and coframe formulations of the torsion gravity field equations.\\
  {In this framework, Noether identities for a special type of quadratic Lagrangian (class $I\!I$ theories, see next section)  has
been worked out  by Nikiforova \cite{nikiforova2017stability}. }

\section{Quadratic Lagrangian\label{QuadLag}}
 The  invariant Lagrangian we shall consider is polynomial, at most of degree two in the curvature, but without  terms explicitly depending only on the torsion~:
 \begin{align}&L_F=  {c_F}\, F+\ft 12 ( f_1\,F_{\alpha\beta}F^{\alpha\beta}+f_2\,F_{\alpha\beta}F^{\beta\alpha})\nonumber\\
 & \phantom{L_F= }+\ft 16(d_1\,F_{\alpha\beta\gamma\delta}F^{\alpha\beta\gamma\delta}+d_2\,F_{\alpha\beta\gamma\delta}F^{\alpha\gamma\beta\delta}+d_3\,F_{\alpha\beta\gamma\delta}F^{\gamma\delta\alpha\beta}) \label{LF}
  \end{align}
 and coupled to the usual Einstein-Hilbert Lagrangian (including a bare cosmological constant)~:
 \begin{align}&L={c_R}\,R-2\,\Lambda + L_F \label{Lagtotquad} 
 \end{align}
 The corresponding tensors needed to write the equations of motion (Eqs [\ref{Dmn}--\ref{STmunu}]) are :
 \begin{align}&Z_\alpha^{{\, .\,}\beta\gamma\delta}=c_F\,\delta_\alpha^{[\gamma}g^{\delta]\beta}+f_1\,F^{\beta[\delta}\delta_\alpha^{\gamma]}+f_2\,\delta_\alpha^{[\gamma}F^{\delta]\beta}+\ft 13\, d_1\,F_\alpha^{{\, .\,}\beta\gamma\delta}-\ft 13\, d_2\,F_{\alpha}^{{\, .\,} [\gamma\delta]\beta}+\ft 13\,d_3\, F^{[\gamma\delta]{\, .\,} \beta}_{\ {\, .\,}{\, .\,} \,\alpha}\qquad ,\\
  &\Delta_{\alpha\beta}=c_F\,F_{(\alpha\beta)}+\ft12\,f_1\,(F_{\alpha\mu}F_\beta^{ {\, .\,}\mu}+F_{  \mu\alpha}F^{ \mu}_{{\, .\,} \beta})+\ft 12\,f_2\,(F_{\alpha\mu}F^{\mu}_{{\, .\,}\beta}+F_{\mu\alpha}F_{ \beta}^{{\, .\,} \mu})\nonumber \\
  &\phantom{\Delta_{\alpha\beta}=}
  +\ft 13\,d_1\,F_{  \mu\nu\rho\alpha}F^{\mu\nu\rho}_{{\, .\,}{\, .\,}{\, .\,} \beta}+\ft 1{6}d_2(F_{\mu\alpha \nu\rho}F_{{\, .\,}{\, .\,}\beta} ^{\mu\nu{\, .\,}\rho}+F_{\mu\beta  \nu\rho}F_{{\, .\,}{\, .\,}\alpha}^{\mu\nu{\, .\,}  \rho}   +F_{ \mu\nu\rho\alpha}F^{\mu \rho\nu}_{{\, .\,}{\, .\,}{\, .\,}\beta} -F_{\alpha\mu\nu\rho}F_\beta^{{\, .\,}\nu\mu\rho} )\nonumber\\
  &\phantom{\Delta_{\alpha\beta}=}+\ft 16 d_3\,(F_{\mu\alpha\nu\rho}F^{\nu\rho\mu}_{{\, .\,}{\, .\,}{\, .\,} \beta}+F_{\mu\beta\nu\rho}F^{\nu\rho\mu}_{{\, .\,}{\, .\,}{\, .\,} \alpha})-\ft 12 g_{\alpha\beta}\,L_F\qquad .
  \end{align}
\\
 Sezgin and van Nieuwenheuizen \cite{PhysRevD.21.3269,PhysRevD.24.1677}, and  Hayashi and Shirafuji  \cite{Hayashi1980gravity1,Hayashi1980gravity2,Hayashi1980gravity3,Hayashi1980gravity4} have  
 analysed a more general nine-parameter Lagrangian obtained by adding to $L_F$ (Eq. [\ref{LF}]) an arbitrary combination of invariant terms quadratic in the torsion tensor components. They have computed the spectrum of the fluctuations their Lagrangian allows around a torsionless flat configuration and established conditions on the parameters that ensure absence of ghosts and tachyonic modes. In general the excitations consist in $0^-$, $0^+$, $1^-$, $1^+$, $2^-$ and $2^+$ fields. In the framework of the models we consider, the $1^\pm$ modes are frozen out and only two classes of field survive. To describe them more precisely let us express the five parameters of the  terms quadratic in the curvature tensor occurring in    Lagrangian $L_F$ (Eq.[\ref{LF}]) in terms of the inverse squared mass of the field fluctuations (labelled by their spin and parity $J^\pm$ ) : $\sigma_{J^\pm}:=2/m^{2}_{J^\pm}$. These parameters are such that the freezeout condition of the mode of spin--parity $J^\pm$ is simply obtained by putting $\sigma_{J^\pm}=0$.
From Ref.  \cite{Hayashi1980gravity4} we obtain :
  \begin{align}&d_1=\frac{c_F}2(\sigma_{2^-}-\sigma_{0^-})\qquad,\label{parad1}\\
 &d_2= {c_F} (\sigma_{2^-}+2\,\sigma_{0^-})\qquad,\\
 &d_3=\frac12\frac{c_F}{c_R}\big(c_R\big(2(\sigma_{2^+}-\sigma_{2^-})+(\sigma_{0^+}-\sigma_{0^-})\big)+c_F\big(2\,\sigma_{2^+}+\sigma_{0^+}\big)\bigg)\qquad,\\
 &f_1=-\frac{c_F}{c_R}\,\frac {( c_R+ c_F)}6\,\big(\sigma_{2^+}+2\,\sigma_{0^+}\big)+\phi\qquad,\\
 &f_2=-\frac{c_F}{c_R}\,\frac {( c_R+ c_F)}6\,\big (\sigma_{2^+}+2\,\sigma_{0^+}\big)-\phi\qquad.\label{paraf2}
\end{align}
 where $\phi$ remains an arbitrary parameter. Of course to avoid tachyons all the $\sigma_{J^\pm}$ have to be non-negative. The fact that all de coupling constant appear to be proportional to $c_F$ results from the expressions of mass  fluctuations (see Eqs [4.11] in Ref.  \cite{Hayashi1980gravity4}) that are all proportional to $c_F$ when the Lagrangian has the form given in Eq. [\ref{Lagtotquad}]. In the limit $c_F=0$ all the masses of the fluctuations vanish and only the $2^-$ modes still contribute to the energy at the quadratic weak field approximation. Of course we may renormalise the $\sigma_{J^\pm}$ in order to maintain all the other coupling constants non zero while the coupling to $F$ is erased, but we prefer to make the assumption that :
 \begin{align}c_F\neq 0\qquad .\end{align}

 The absence of ghosts  restricts much more the possible configurations, leading to   two classes of physically acceptable Lagrangian of the type Eq. [\ref{Lagtotquad}].
 The first one, usually discarded, contains in addition to the massless spin 2 modes, only massive $0^+$ and $2^-$ modes. It is characterised 
by the parameter restrictions~:
\begin{align}\text{Class $I$ :}\qquad c_R\geq 0\quad,\quad c_F< 0\quad,\quad \sigma_{0^-}=\sigma_{2^+}=0\quad,\quad\sigma_{0^+}\geq0\qquad,\qquad\sigma_{2^-}\geq0\qquad.\label{cl1}
\end{align}
 The second one contains as massive modes only $0^-$ and $2^+$ fields. It requires that :
 \begin{align}\text{Class $I\! I$ :}\qquad c_R\geq 0 \quad,\quad c_F> 0\quad,\quad\sigma_{0^+}=\sigma_{2^-}=0\quad,\quad\sigma_{0^-}\geq0\qquad,\qquad\sigma_{2^+}\geq0\qquad.\label{cl2}
 \end{align}
Let us recall that in order to recover in the usual coupling of the massless spin 2 field we have to impose :
 \begin{align}c_R+c_F=  \frac 1{16\,\pi\,G}=:\kappa_N
 \end{align}
 where $G$ is Newton's constant.
 
  These relations ensure that Eq. [\ref{Lagtotquad}] provide the most general invariant Lagrangian depending only on the curvature tensor, at most quadratic in it, and  physically acceptable ({i.e.} without ghost or tachyon at the level of quadratic fluctuations around flat space, in absence of background torsion).  
  \section{Special solutions}
  { From now we assume the Lagrangian given by Eq. [\ref{Lagtotquad}], possibly  restricted by constraints on the coupling parameters : $f_1,\dots,\, d_3$ but with $c_R\neq 0$ and $c_F\neq 0$.
 Various aspects of exact or numerical solutions of the class $I\!I$ torsion gravity field equations it provides have been discussed in the literature (see for instance Refs \cite{PhysRevD.80.104031,PhysRevD.95.024013,nikiforova2017stability,Nikiforova_2018,damour2019spherically,nikiforova2020black}).  Here after we shall display some exact solutions of models of classes $I$ and $I\! I$, mainly in order to illustrate the differences between the two. We start by briefly considering torsionless solutions then turn to torsionful solutions.}
 \subsection{Torsionless solutions}
 First let us assume that the contorsion   vanishes. 
 In order to simplify notations, we denote~:
\begin{align}&\ft 12(f_1+f_2)=-\frac{c_F}{c_R}\frac{\kappa_N}6\,(2\,\sigma_{0^+}+ \sigma_{2^+})=:\bar f\qquad,\\
& {\ft 13(d_1+\ft 12\,d_2+d_3)} =\frac{c_F}{c_R}\frac{\kappa_N}6\,(\sigma_{0^+}+2\,\sigma_{2^+})=:\bar d\qquad .
\end{align}
The quadratic Lagrangian and the various tensors appearing in the field equations reduce to~:
\begin{align}&L_F={c_F}\,R+\bar f\,R_{\mu\nu}R^{\mu\nu}+\ft 12\,{\bar d}\,R_{\mu\nu\rho\sigma}R^{\mu\nu\rho\sigma} \qquad ,\\
&Z^{\alpha\beta\gamma\delta}=c_F\,g^{\alpha[\gamma}g^{\delta]\beta}+2\,\bar f\,g^{\alpha[\gamma}R^{\delta]\beta}+ {\bar d}\,R^{\alpha\beta\gamma\delta} \qquad ,\\
&F_{(\alpha \mu\nu\rho}\,S_{\beta)}^{\,{\, .\,} \mu\nu\rho}=\bar f\,(R_{\alpha\mu\beta\nu}\,R^{\mu\nu}-R^\mu_\alpha\,R_{\mu\beta}) \qquad ,\\
&\Delta_{\alpha\beta}={c_F}(R_{\alpha\beta}-\ft 12\,g_{\alpha\beta} \,R)+2\,\bar f\,(R_\alpha^\mu\,R_{\mu\beta}-\ft14\,g_{\alpha\beta}\,R_{\mu\nu}R^{\mu\nu})\nonumber\\
&\phantom{\Delta_{\alpha\beta}=}+\bar d\,(R_{\alpha\rho\mu\nu}R_\beta^{{\, .\,} \rho\mu\nu}-\ft 14\,g_{\alpha\beta}\,R_{\mu\nu\rho\sigma}R^{\mu\nu\rho\sigma}) \label{Del1}\\
&\phantom{\Delta_{\alpha\beta}}={c_F}\big(R_{\alpha\beta}-\ft 12\,g_{\alpha\beta} \,R)+2\,\bar f\,\big(R_\alpha^\mu\,R_{\mu\beta}-\ft14\,g_{\alpha\beta}\,R_{\mu\nu}R^{\mu\nu}) \nonumber \\
&\phantom{\Delta_{\alpha\beta}=}+\bar d\,\big(2\,R_{\alpha\mu\beta\nu}R^{\mu\nu}+2\,R_{\alpha\mu}R^\mu_\beta- R\,R_{\alpha\beta}-g_{\alpha\beta} (R_{\mu\nu}R^{\mu\nu}-\ft14\,R^2)\big) \label{Del2}\qquad .
\end{align}
As in Ref.  \cite{Hayashi1980gravity4}, the writing of the last equation is simplified by use of the Bach-Lanczos \cite{Bach:1921aa,Lanczos1938} identity (see Appendix [\ref{GBE}] for a topology based proof of it).

  { The connection field equations  reduce to~:
 \begin{align}{\nabla\!}_\nu Z^{[\alpha\beta]\gamma\nu}=0\qquad ,
 \end{align}
 {i.e. } to~:
 \begin{align}c_F\big(( 2\,\sigma_{0^+}+7\,\sigma_{2^+}){\nabla\!}_{[\alpha }R_{\beta]\gamma} +( 2\,\sigma_{0^+}+ \sigma_{2^+}) \,g_{\gamma[\alpha }{\nabla\!}_{\beta ]}R\big)=0 \qquad ; \label{KeqKnul}
 \end{align}
 and the Einstein equations become~:
 \begin{align}& c_R\Big(R_{\alpha\beta}-\ft 12\,g_{\alpha\beta}\,R+\frac{\Lambda}{\kappa_N}\,g_{\alpha\beta}\Big)
 +   {c_F} \Big(\frac{\sigma_{2^+}}2 C_{\alpha\mu\beta\nu}R^{\mu\nu}  
  -\frac{ \sigma_{0^+} }6\,R\,{\big(}R_{\alpha\beta}-\ft 14\,g_{\alpha\beta}\,R\big)\Big)=0\label{EeqKnul}
  \end{align}
 where $C_{\alpha \beta\gamma\delta}$ are the components of the Weyl tensor.\newline
 Unless $\sigma_{0^+}$ and $\sigma_{2^+}$ vanish (but recall that the absence of ghosts and tachyons imply that both cannot be simultaneously non-zero and that $c_F$ is assumed to be non-zero), Eqs [\ref{KeqKnul}] implies   that the scalar curvature is constant ~:
 \begin{align}
 R=\frac 4{\kappa_N}\,\Lambda \label{R4L}
 \end{align}
 and the field equations [\ref{KeqKnul}, \ref{EeqKnul}] are equivalent to :
\begin{align}&{\nabla\!}_{[\alpha }R_{\beta]\gamma}=0\qquad ,\label{DRA}\\
&(c_R+\frac 23\,c_F\,\sigma_{0^+}\frac \Lambda{\kappa_N})(R_{\alpha\beta}-\frac \Lambda{\kappa_N}\,g_{\alpha\beta})=\frac 12\,c_F\,\sigma_{2^+}\, C_{\alpha\mu\beta\nu}R^{\mu\nu} \qquad .\label{RCR}
\end{align}

We now note the following~:
\begin{itemize}
\item In case $\sigma_{0^+}=0$ and $\sigma_{2^+}=0$, {i.e. } for models including only odd-parity fields (of spin two in the framework of  class {$I$} models, of spin zero for those of  class {$I\! I$}) , if we assume a vanishing torsion, the field equations reduce to the usual standard Einstein equations.
\item More generally, based on the method developed in the seminal work of Debney {\it et al}.  \cite{debney1978equivalence}, Obukhov {\it et al}. \cite{obukhov1989quadratic}  have proved that this conclusion remains valid in the generic case : only Einstein spaces metrics are solutions of  the Eqs [\ref{R4L} -- \ref{RCR}]   when $\kappa_N\,c_R/(c_F\,\Lambda)\neq -3/2\{\sigma_{0^+},\sigma_{0^+}+\ft12\,\sigma_{2^+},\sigma_{0^+}-\sigma_{2^+}\}$.
\item If we assume that the metric is conformally flat, Eq.  [\ref{RCR}]  implies,   when $1-\ft 23\,\sigma_{0^+}\,(c_F/c_R)\,(\Lambda/\kappa_N)\neq 0$,  that the space is a conformally flat Einstein space, {\i.e.} a flat, de Sitter or anti-de Sitter space.
\item To explore the special case $\Lambda=\ft 32\,\kappa_N\,c_R/(c_F\,\sigma_{0^+})$ let us  assume the  (conformally flat) metric to be 
\begin{align}ds^2=-dt^2+Y(t)\Big(\frac{dr^2}{1+k\,r^2}+r^2\,d\theta^2+r^2\,\sin^2(\theta)\,d\varphi^2 \Big)\qquad\label{spmet}
\end{align}
with $k=0$ or $k=\pm L^{-2}$, $L$ being a constant and $Y(t)$ a positive function. \\ Then we obtain :
\begin{itemize}
\item For class {$I$} models (Eqs [3.11]) the trace of the Einstein equations leads to  the second order differential equation :
\begin{align}   \ddot Y(t)-\frac{2\,c_R}{c_F\,\sigma_{0^+}}\,Y(t)=2\,k
\end{align}
whose solution reads after having fixed appropriately the origin of the $t$ coordinate (and assuming $c_R\neq 0$ ) :
\begin{align}
Y(t)=\left\{\begin{array}{lll}
-k\,\frac {c_F}{c_R}\,\sigma_{0^+}+\frac 12\,Y_0\bigg(e^{ \sqrt{\ft{2\,c_R}{c_F\,\sigma_{0^+}}}\,t}+\varepsilon \,e^{-\sqrt{\ft{2\,c_R}{c_F\,\sigma_{0^+}}}\,t}\bigg)\quad&, \quad\varepsilon=\pm 1\qquad &\text{if }\Lambda >0\qquad,\\
-k\,\frac {c_F}{c_R}\,\sigma_{0^+}+ Y_0\,\cos\Big(\sqrt{-\ft{2\,c_R}{c_F\,\sigma_{0^+}}}t\Big) &, \quad\varepsilon=+ 1\qquad &\text{if }\Lambda <0\qquad,
\end{array}\right .\label{Yt}
\end{align}
  $Y_0$ being an integration constant whose sign must be chosen to insure (at least for some values of $t$) the positivity of $Y(t)$.\\
Inserting the expression of the metric [\ref{spmet}] obtained from Eq. [\ref{Yt}] in the remaining field equations we see that they are all satisfied for the special value of $\Lambda$ here considered.\\
The components of the Ricci tensor of the corresponding metrics are :
\begin{align} R^t_t&=\frac{3\, c_R}{2\,c_F\,\sigma_{0^+}}\Bigg(1-\frac{ (c_F^2\,\sigma_{0^+}^2\,k^2-\varepsilon \,c_R^2\,Y_0^2)}{c_R^2\,Y^2(t)}\Bigg)\qquad,\\
 R^r_r=R^\theta_\theta=R^\varphi_\varphi&=\frac{3\, c_R}{2\,c_F\,\sigma_{0^+}}\Bigg(1+\frac{ (c_F^2\,\sigma_{0^+}^2\,k^2-\varepsilon \,c_R^2\,Y_0^2)}{3\,c_R^2\,Y^2(t)}\Bigg)\qquad.
\end{align}
In particular, for $\varepsilon=+1$, when :
\begin{align}Y_0^2=\Big(\frac{c_F\,\sigma_{0^+}\,k}{ c_R}\Big)^2\qquad .\label{dScond}
\end{align}
the geometries are those of anti-de Sitter or de Sitter spaces.\\ If this condition is not fulfilled, the behaviour of the trace of the square of the Ricci tensor :
\begin{align}R^\alpha_\beta\, R^\beta_\alpha=\frac {9\,c_R^2}{c_F^2\,\sigma_{0^+}}+3\,\frac{ (c_F^2\,\sigma_{0^+}^2\,k^2-\varepsilon\, c_R^2\,Y_0^2)^2}{c_F^2\,\sigma_{0^+}^2\,c_R^2\,Y^4(t)}
\end{align}
 shows that the vanishing of $Y(t)$ corresponds to a true curvature singularity and the solution describes a space evolving between two cosmological singularities.\\
It is interesting to note that all these solutions  are independent of the coupling constant $\sigma_{2^-} $.
\end{itemize}
\item  Notice that the Lagrangian of Yang's theory \cite{yang1974integral} only involves a term proportional to $F_{\alpha\beta\gamma\delta}F^{\alpha\beta\gamma\delta}$. Accordingly, torsionless connections have to be solutions of Eq. (\ref{DRA})  and to satisfy the condition :
\begin{align}
R_{\alpha\mu\nu\rho}R^{\beta\mu\nu\rho}-\frac 14\,\delta_\alpha^\beta \,R_{ \mu\nu\rho\sigma}R^{\mu\nu\rho\sigma}=0
\end{align}
instead of Eq. (\ref{RCR}). Some examples of such solutions are displayed in Refs \cite{ni1975yang,pavelle1975unphysical,pavelle1976unphysical}, but in the framework of dynamical torsion gravity, even with an Einstein-Hilbert piece added to it, we have to emphasise that this Lagrangian is not gosht-free or tachyon-free.
\end{itemize}
}

\subsection{Torsionfull solutions}
In this section we will integrate the field equations under appropriate simplifying ansatzes. We start from a prescribed form of the metric which solves the field in absence of torsion and deform it by introducing  a minimal contorsion that maintains the equations tractable. We shall consider four different kinds of metrics : Bertotti-Robinson, pp-wave, Friedmann-Lema\^itre-Robertson-Walker, black hole. Each of them constitutes a solution of the field equation in the absence of torsion. The first one constitute the most surprising one : the resulting torsion depends on arbitrary functions that are completely ignored by the metric. A similar property also appears for the black hole configuration.
\subsubsection{Bertotti-Robinson geometry}
Equation [\ref{DRA}] shows that the metric of symmetric spaces are solutions of the contorsion equations when it vanishes. Among such symmetric spaces \cite{cahen1968metriques}, let us first consider the  Bertotti-Robinson space\cite{robinson1959solution,bertotti1959uniform}. This space is the Riemannian product of an Euclidian and a Lorentzian space : the product of a sphere or a hyperbolic plane\footnote{We will not discuss possible compactification obtained by quotienting by a Fushian group, nor other global aspects of the geometry.}  and a Lorentzian factor : a de Sitter or anti-de Sitter bidimensionnal space. In local coordinates   $\{r,\,\theta,\,\xi,\,\tau\}$    the metric tensor reads~:
\begin{align}ds^2=\frac 1{(1+Q_1\,r^2)^2}(dr^2+r^2\,d\theta^2)+\frac 1{(1+Q_2\,\xi^2)^2}(d\xi^2-\xi^2\,d\tau^2)\qquad.\label{RobBertmet}
\end{align}
It is an Einstein space when $Q_1=Q_2$  {, a conformally flat space when $Q_1=-Q_2$ }.\\ 
To pursue we made the following ansatz for the expression of the contorsion tensor components~:
\begin{align}&K_{r\theta r}=-K_{\theta r r}:= \ft{1}{(1+Q_1\,r^2)}k_\theta(r) \quad,\quad K_{\theta r \theta}:= \ft{r^2}{(1+Q_1\,r^2)}k_r(r)=-K_{r\theta \theta}\qquad,\qquad \nonumber\\
&K_{\xi\tau\xi}:= \ft{1}{(1+Q_2\,\xi^2)}k_\xi(\xi)=-K_{\tau \xi\xi}\qquad,\qquad K_{\tau\xi\tau}:=- \ft{\xi^2}{(1+Q_2\,\xi^2)}k_\tau(\xi)=-K_{\xi\tau\tau}\qquad,\label{RBKassump}
\end{align}
all the other components being assumed to be zero.\\
The ten non-vanishing contorsion field equations depend only on two terms~:
\begin{align}{\kappa_E}(r):=(1+Q_1\,r^2)^2\Big(k'_r(r)+\frac 1r\,k_r(r)\Big)\qquad,\qquad {\kappa_L}(\xi)=(1+Q_2\,\xi^2)^2\Big(k'_\xi(\xi)+\frac 1\xi\,k_\xi(\xi)\Big)\qquad .\label{kEkLPol}
\end{align}
They read as~:
\begin{align} &{{\mathcal S}}^\tau_{{\, .\,}\theta\tau}={{\mathcal S}}^\xi_{{\, .\,}\theta\xi}  
 = k_\theta(r){\big(}c_F+2{\,\bar f\,}(Q_1+Q_2)-\ft 12{\,\bar f\,}({\kappa_E}(r)+{\kappa_L}(\xi))\big)=0\quad ,\\
&{{\mathcal S}}^\theta_{{\, .\,}\tau\theta}={{\mathcal S}}^r_{{\, .\,}\tau r}= k_\tau(\xi){\big(} c_F+2{\,\bar f\,}(Q_1+Q_2)-\ft 12{\,\bar f\,}({\kappa_E}(r)+{\kappa_L}(\xi))=0\quad ,\\
&{{\mathcal S}}^\theta_{{\, .\,} r\theta}=-{\big(}\ft 23\,\bar d\,+{\,\bar f\,}\big){\kappa_E}'(r)=0\quad ,\\
&{{\mathcal S}}^\xi_{{\, .\,}\tau\tau}=-{\big(}\ft 23\,\bar d\,+{\,\bar f\,}\big){\kappa_L}'(\xi)=0\quad ,\\
&{{\mathcal S}}^\tau_{{\, .\,} r\tau}={{\mathcal S}}^\xi_{{\, .\,} r\xi}=- \ft 12{\,\bar f\,}{\kappa_E}'(r)\nonumber\\
&\phantom{{{\mathcal S}}^\tau_{{\, .\,} r\tau}={{\mathcal S}}^\xi_{{\, .\,} r\xi}=}-k_r(r){\big(}c_F+2{\,\bar f\,}(Q_1+Q_2)- \ft 12{\,\bar f\,}({\kappa_E}(r)+{\kappa_L}(\xi))\big)=0\quad ,\\
&{{\mathcal S}}^r_{{\, .\,} \xi r}={{\mathcal S}}^\theta_{{\, .\,} \xi \theta}=- \ft 12{\,\bar f\,}{\kappa_L}'(\xi)\nonumber\\
&\phantom{{{\mathcal S}}^r_{{\, .\,} \xi r}={{\mathcal S}}^\theta_{{\, .\,} \xi \theta}=}-k_\xi(\xi){\big(}c_F+2{\,\bar f\,}(Q_1+Q_2)- \ft 12{\,\bar f\,}({\kappa_E}(r)+{\kappa_L}(\xi))\big)=0\quad .
\end{align}
Their solutions are given by ~:
\begin{align}{\kappa_E}(r)=2(Q_1+Q_2)+\frac{ c_F}{{\,\bar f\,}}+\beta\qquad,\qquad {\kappa_L}(\xi)=2(Q_1+Q_2)+\frac{c_F}{{\,\bar f\,}}-\beta\label{kEkLPolsol}
\end{align}
where $\beta $ is a constant.
From them we obtain~:
\begin{align}&k_r(r)=\frac{\alpha_r}r- \frac{c_F/{\,\bar f\,}+2(Q_1+Q_2)+\beta}{2\,Q_1\,r\,(1+Q_1\,r^2)}\quad,\label{RBkr}\\
& k_\xi(\xi)=\frac{\alpha_\xi}{\xi}- \frac{c_F/{\,\bar f\,}+2(Q_1+Q_2)-\beta}{2\,Q_2\,\xi\,(1+Q_2\,\xi^2)}\label{RBkxi}
\end{align}
with $\alpha_r$ and $\alpha_\xi$ two integration constants.\\
Using these expression of $k_r(r)$ and $k_\xi(\xi)$, the four non trivial remaining Einstein equations reduce to two algebraic equations~:
\begin{align}&\mathcal E^r_r=\mathcal E^\theta_\theta=\Lambda-4\,c_R\,Q_2-\frac{4\,c_F^2\,{\bar d}}{3{\,\bar f\,}^2}+ \frac{4\,c_F(\beta- 4\,Q_1){\big(}\bar d+\ft34{\,\bar f\,}\big)}{3{\,\bar f\,}}=0\qquad,\\
&\mathcal E^\xi_\xi=\mathcal E^\tau_\tau=\Lambda-4\,c_R\,Q_1+\frac{4\,c_F^2\,({\bar d}+\ft32{\,\bar f\,})}{3{\,\bar f\,}^2}- \frac{4\,c_F(\beta- 4\,Q_1){\big(}\bar d+\ft34{\,\bar f\,}\big)}{3{\,\bar f\,}}=0\qquad,
\end{align}
that fix the value of the curvatures of the two factors of the product geometry~:
\begin{align}&Q_1=\frac{4\,c_F({\bar d}+\ft 34 {\,\bar f\,}){\big(}c_F^2+{\,\bar f\,}(\beta\,c_R+\Lambda)\big)-3\,c_R{\,\bar f\,}{\big(}c_F^2+{\,\bar f\,}\Lambda\big)}{16\,c_R{\,\bar f\,}(c_F({\,\bar d}+\ft 34 {\,\bar f\,})-\ft 34\,c_R{\,\bar f\,}}\\
&Q_2=\frac{4\,c_F({\,\bar d}+\ft 34 {\,\bar f\,}){\big(}c_F^2-{\,\bar f\,}(\beta\,c_R-\Lambda) \big)-3\,c_R{\,\bar f\,}{\big(}c_F^2+{\,\bar f\,}\Lambda\big)}{16\,c_R{\,\bar f\,}(c_F({\,\bar d}+\ft 34 {\,\bar f\,})-\ft 34\,c_R{\,\bar f\,}}
\end{align}\\
The solution here above involves two arbitrary functions of one variable : $k_\theta(r),\ k_\tau(\xi)$ and three integration constants : $\alpha_r$, $\alpha_\xi$ and $\beta$. \\

The previous solutions could be slightly generalised. The coordinate $\theta$ used to write the metric Eq. [\ref{RobBertmet}] is an angle. Accordingly functions depending on it have to be periodic. But the time coordinate $\tau$ varies from $-\infty$ to $+\infty$ and functions depending on it have no {  {\it a priori}  } restriction. This suggests to consider the arbitrary functions appearing in the $K_{r\theta r}$ and $K_{\theta r \theta}$ components of the contorsion to depend also on $\tau$.\\
Written as~:
\begin{align}&K_{r\theta r}=-K_{\theta r r}:= \ft{1}{(1+Q_1\,r^2)}\tilde k_\theta(r,\tau) \quad,\quad K_{\xi\tau\xi}=-K_{ \tau\xi\xi}:= \ft{1}{(1+Q_2\,\xi^2)}\tilde k_\xi(\xi,\tau)\qquad ,
\end{align}
the $(\xi,\ \theta,\ \tau)$ contorsion equation  
\begin{align}{\mathcal S}^\xi_{{\, .\,} \theta\tau}=\frac {f_1(Q_2^2\,\xi^4-1)}{2\,\xi}\partial_\tau \tilde k_\theta(r,\tau)=0 \end{align}
implies, assuming to be in a generic case, that $K_{r\theta r}$ cannot depends on $\tau$, but is an arbitrary function of $r$ :
\begin{align}\tilde k_\theta(r,\tau)=k_\theta(r)\qquad.\end{align}
 On the other hand the $(\xi,\ \tau,\ \xi)$ equation  
\begin{align}&{\mathcal S}^\xi_{{\, .\,} \tau\xi}=-\frac{({\,\bar d}+\ft 32{\,\bar f\,})(1+Q_2\,\xi^2)^6}{3\,\xi^4}\,\partial^2_{\tau\tau}\tilde k_\tau(\xi,\tau)=0
\end{align}
implies that $k_\tau(\xi,\tau)$ is linear in the $\tau$ coordinate. Under these assumptions we obtain a solution with $k_r(r)$ still given by Eq. [\ref{RBkr}], depending on two integration constants but with $k_\xi(\xi)$ an arbitrary function and $k_\tau(\xi,\tau)$ given by
\begin{align}& \tilde k_\tau(\xi,\tau)=k_\tau(\xi)+\Big(\xi\,k_\xi(\xi)+\xi^2\,{\big(}k_\xi'(\xi)-\frac{(2\,c_F/{\,\bar f\,}+2\,(Q_1+Q_2)-\beta)}{(1+Q_2\,\xi^2)^2}\big)\Big)\tau\end{align}
where $k_\tau(\xi)$ is another arbitrary function.\\
More involved solution may be obtained by performing a coordinate transformation from polar coordinates  to planar coordinates. For illustrative purpose let us suppose $Q_2=4\,L^2>0$ and the metric
given by :
\begin{align}ds^2=\frac 1{(1+Q_1\,r^2)^2}(dr^2+r^2\,d\theta^2)+\frac 1{4\,Q_2\,t^2}(dz^2-dt^2)\end{align}
The new planar coordinates $(t,\,z)$ are related to the polar coordinates by :
\begin{align}t=\frac {L(4\,L^2+\xi^2)}{4\,L^2-\xi^2-4\,L\,\sinh (\tau)}\qquad,\qquad z=\frac { 4\,L^2\,\xi\,\cosh(\tau)}{4\,L^2-\xi^2-4\,L\,\sinh (\tau)}\qquad .\end{align}
On this new coordinate patch the non-vanishing contorsion components are, in accordance with the ansatz [\ref{RBKassump}]~: 
\begin{align}&K_{r\theta r}=-K_{\theta r r}:= \ft{1}{(1+Q_1\,r^2)}k_\theta(r) \qquad,\qquad K_{\theta r \theta}=-K_{r\theta \theta}:= \ft{r^2}{(1+Q_1\,r^2)}k_r(r)\qquad,  \nonumber\\
&K_{ztz}=-K_{tzz}:= \ft{1}{(4\,Q_2\,t^2)}k_t(t,z)\qquad,\qquad K_{tzt}=-K_{ztt}:=-\ft{1}{(4\,Q_2\,t^2)}k_z(t,z) \qquad .\end{align}
Again the contorsion equations depend only on two terms : $\kappa_E(r)$ already defined in Eq. [\ref{kEkLPol}] and 
\begin{align}\tilde\kappa_L(t,z):=4\,Q_2\,t^2\big(\partial_zk_z(t,z)-\partial_tk_t(t,z)\big)\qquad .\end{align}
The structure of the system of equations remains the same. The  equations that $\kappa_E(r)$ and $\tilde\kappa_L(t,z)$ have to satisfy are similar to Eqs [\ref{kEkLPolsol}] :
\begin{align}{\kappa_E}(r)=2(Q_1+Q_2)+\frac{ c_F}{{\,\bar f\,}}+\beta\qquad,\qquad {\tilde\kappa_L}(t,z)=2(Q_1+Q_2)+\frac{c_F}{{\,\bar f\,}}-\beta\label{kEkLPlasol}\qquad .\end{align}
Accordingly, the function $k_\theta(r)$ remains an arbitrary function of one variable, $k_r(r)$ is still given by Eq.[\ref{RBkr}] but :
\begin{align}k_z(t,z)=A_z(t,z)\qquad k_t(t,z)=\int\partial_zA_z(t,z)\,dt+b_t(z)+\frac{c_F/{\,\bar f\,}+2(Q_1+Q_2)-\beta}{4\,Q_2\,t^2}\qquad.\label{Atz}\end{align}
Using the Jacobian defined by the coordinate transformation ($k_z,\ k_t$ transforms as a covector components), we easily obtain a solution in polar coordinate that generalise the previous one.

To take into account the conditions   that ensure the health of the theory around flat space we rewrite the solution using the parametrisation (Eqs[\ref{parad1}--\ref{paraf2}])  :
\begin{align} &Q_1= \frac{1}{4} \left(\frac{ \Lambda }{c_R}+\frac{3   
  {\,c_F }{\,\sigma_{2^+}\,}\beta}{3{\,c_F }{\,\sigma_{2^+} }+2{\,c_R }
  {\,\sigma_{0^+} }+c_R{\,\sigma_{2^+} }}-\frac{3
  {\,c_F }}{ \kappa_N\, (2\,
  {\sigma_{0^+} }+\sigma_{2^+})}\right)\qquad,\\
   &Q_2= \frac{1}{4} \left(\frac{ \Lambda }{c_R}-\frac{3   
  {\,c_F }{\,\sigma_{2^+}\,}\beta}{3{\,c_F }{\,\sigma_{2^+} }+2{\,c_R }
  {\,\sigma_{0^+} }+c_R{\,\sigma_{2^+} }}-\frac{3  {\,c_F }}{ \kappa_N\,(2\,
  {\sigma_{0^+} }+\sigma_{2^+})}\right) \qquad .
 \end{align}
  Notice that \
  \begin{align}\bar f=-\frac {\kappa_N}{6}\,\frac{c_F}{c_R}\,\big(\sigma_{2^+}+2\,\sigma_{0^+}\big)\qquad .
  \end{align}
  Thus the Bertotti-Robinson geometry, for the contorsion ansatz here assumed, is only compatible, in the framework of a class {$I$} model  with a $0^+$ field, and for a class {$I\! I$}  model with a $2^+$ field. It is easy to check that if $\sigma_{0^+}$ and $\sigma_{2^+}$ vanish no torsionful solutions are possible in the framework here considered.  { Notice that, on the contrary to what is done in Ref. \cite{zhytnikov1993conformally}  we did not assume the metric to be conformally flat and the contorsion field we obtain as solution of the field equations depends on arbitrary functions instead of arbitrary constants.}
  \subsubsection{Plane-fronted wave spacetimes}
  This family of spaces is known since a long time \cite{Brinkmann} and has been extensively studied, in particular their physical interest being put into evidence in numerous works (See for instance Refs  \cite{ehlers1962exact,Penrose1965,penrose1976any,deser1975plane}). Among them there also is a subset of symmetric Lorentzian spaces \cite{cahen1968metriques}, making them exact torsionless solutions of the torsion gravity equations. \\
  The general expression of their metric is :
  \begin{align}ds^2=2\,du\,dv+H(u,x,y)\,du^2+dx^2+dy^2\label{ppwmet}
  \end{align}
  In absence of contorsion it solves the field equations (both for classes I and II models) if and only if $H(u,x,y)$ is an harmonic function with respect to the $x$ and $y$ coordinates and the bare cosmological constant vanishes :
  \begin{align}\Lambda=0\qquad,\qquad \Delta \,H(u,x,y)=0\qquad .\end{align}
 Here $\Delta$ denotes the   two dimensionnal  Laplacian ($\Delta := {\partial_x^2} + {\partial_y^2}$).\\
    {Extension of this solution to torsionful configurations has been discussed with great generality in Ref. \cite{adamowicz1980plane}. To display some explicit solutions we assume as {  {\it a priori}  } non-vanishing components of the contorsion :}
  \begin{align}K_{xuu}=-K_{uxu}=:X(u,x,y)\qquad,\qquad K_{yuu}=-K_{uyu}=:Y(u,x,y)\qquad .\end{align}
  We will now briefly discuss the resolution of the field equations under this ansatz :
   \begin{itemize}
 \item   {Class $I$ models : Einstein equations continue to impose to put $\Lambda =0$. Nevertheless let us mention that solutions representing pp-wave propagating on (anti-) de Sitter spaces are described in Ref. \cite{blagojevic2017generalized}).}Two independents contorsion equations ($\mathcal S^x_{{\, .\,} vv}=0$, $\mathcal S^y_{{\, .\,} vv}=0$) and two independent  Einstein equations ($\mathcal E^u_v=0$, $\mathcal E^u_u=0$) remain to be solved.
  We obtain from the contorsion equations :
  \begin{align}&c_F\big( {\partial_x^2}  Y(u,x,y)- {\partial^2_{xy}}  X(u,x,y)-\frac{ 2} {\sigma_{2^-}}Y(u,x,y)\big)=0\\
&c_F\big( {\partial_y^2}  X(u,x,y)- {\partial^2_{xy}}  Y(u,x,y)-\frac{ 2} {\sigma_{2^-}}X(u,x,y)\big)=0 \end{align}
 from which we deduce that (let us recall that we assume $c_F\neq 0$) :
 \begin{align}\partial_x X(u,x,y)+\partial_y Y(u,x,y)=0 \label{dXdY}\end{align}
 {i.e. }  the contorsion components $K_{xuu}$ and $K_{yuu}$ have to verify  a two dimensional Helmholtz  equation :
 \begin{align}\Delta \,X(u,x,y)-\frac {2}{\sigma_{2^-}}\,X(u,x,y)=0=\Delta \,Y(u,x,y)-\frac {2}{\sigma_{2^-}}\,Y(u,x,y)\qquad .\end{align}
 Accordingly  $X(u,x,y)$ ({\it resp}.$Y(u,x,y)$) may be written as a superposition of exponential modes $\xi_I (u,x,y)$ ({\it resp}.$ \eta_I (u,x,y)$) : 
 \begin{align}& \xi_I (u,x,y)= \pm \sqrt{\Big(\frac 2{\sigma_{2^-}}-p^2(u)\Big)}\,k(u)\,e^{-p(u)\,x\pm\sqrt{\frac 2{\sigma_{2^-}}-p^2(u)}\,y}\qquad,\\
 & \eta_I (u,x,y)=   p(u)\,k(u)\,e^{-p(u)\,x\pm\sqrt{\frac 2{\sigma_{2^-}}\pm p^2(u)}\,y}\qquad ,\end{align}
 $p(u)$ and $k(u)$ being arbitrary functions of $u$, only limited by the condition that the contorsion components have to be real.\\
 The diagonal Einstein equations impose that the bare cosmological constant $\Lambda$ vanishes. Then, taking this into account and the relation Eq.(\ref{dXdY}), we obtain from the remaining equation ~:
 \begin{align}\Delta H(u,x,y)=0\label{ppwvvacsol}\end{align}
 {i.e. } the geometry is still Ricci flat.  Notice that as $x$ and $y$ varies from $-\infty$ to $+\infty$  we are confronted by an exponential blowup of the contorsion (the metric blowing up only polynomially).
 \item Class  $I\! I$ models : In this case the equations apparently look a little bit more complicated but lead to a similar solution. We have still to impose $\Lambda =0$. The contorsion equations read 
  \begin{align}&c_F\Big( {\partial_x^2}  X(u,x,y)+ {\partial^2_{xy}}  Y(u,x,y)-\frac{ 2\,c_R} {\kappa_N\,\sigma_{2^+}}X(u,x,y)-\frac 12\,\partial_x\Delta H(u,x,y)\Big)=0\label{eqppwX}\qquad ,\\
&c_F\Big( {\partial_y^2}  Y(u,x,y)+ {\partial^2_{xy}}  X(u,x,y)-\frac{ 2\,c_R} {\kappa_N\,\sigma_{2^+}}Y(u,x,y)-\frac 12\,\partial_y\Delta H(u,x,y)\Big)=0 \label{eqppwY}\qquad ,
\end{align}
 from which we obtain :
 \begin{align}\partial_yX(u,x,y)=\partial_x Y(u,x,y)\qquad .\end{align}
 The diagonal Einstein equations still imply the vanishing of the bare cosmological constant. The remaining non trivial equation is :
 \begin{align}\Delta H(u,x,y)=2\frac{c_F}{\kappa_N}\big(\partial_yY(u,x,y)+\partial_xX(u,v,y)\big)\label{eqppHcl2}
 \end{align}
 which inserted in Eqs [\ref{eqppwX},\ref{eqppwY}] leads again to  Helmholtz equations :
\begin{align}\Delta \,X(u,x,y)-\frac {2}{\sigma_{2^+}}\,X(u,x,y)=0=\Delta \,Y(u,x,y)-\frac {2}{\sigma_{2^+}}\,Y(u,x,y) \qquad.\end{align}
From these last two   we deduce that the solution of Eq. [\ref{eqppHcl2}] is given by :
\begin{align}H(u,x,y)=H_0(u,x,y)+\frac{c_F\,\sigma_{2^+}}{\kappa_N}\big(\partial_yY(u,x,y)+\partial_xX(u,x,y)\big)\end{align}
$H_0(u,x,y)$ being an harmonic function and the contorsion components $X(u,x,y)$ and $Y(u,x,y)$ by   superpositions of modes :
\begin{align}& \xi_{II} (u,x,y)=  p(u)\,k(u)\,e^{-p(u)\,x\pm \sqrt{\frac 2{\sigma_{2^-}}-p^2(u)}\,y}\qquad,\\
 & \eta_{II} (u,x,y)=\mp \sqrt{\frac 2{\sigma_{2^-}}-p^2(u)}\,k(u)\,e^{-p(u)\,x\pm\sqrt{\frac 2{\sigma_{2^-}}-p^2(u)}\,y}\qquad ,\end{align}
   $p(u)$ and $k(u)$ being arbitrary functions of $u$, satisfying the same conditions as those encountered for class {$I$} pp-wave solutions. Let us emphasise that the function $H(u,x,y)$ being no more an harmonic function (in $x$ and $y$), on the contrary to the metric obtain in the framework of class $I$ pp-wave, the metric no longer constitutes a solution of a vacuum Einstein space.
 \end{itemize}
  \subsubsection{Friedmann-Lema\^itre-Robertson-Walker spatially flat geometry\label{FLRW}}
  The simplest cosmological  geometry is  :
  \begin{align}ds^2=-dt^2+e^{2\,A(t)}(dx^2+dy^2+dz^2) \end{align}
  describing a spatially flat homogeneous and isotropic space.
  In this section we discuss some solutions of torsion gravity theories under the assumptions that the metric is such one and that the contorsion tensor is of the special form :
  \begin{align}K_{\alpha\beta\gamma}=\eta_{\alpha\beta\gamma\delta}a^\delta +\frac 13\,(g_{\alpha\gamma}\,k_{\beta} -g_{\beta\gamma}\,k_{\alpha})\label{FLRWK}\end{align}
  with 
  \begin{align}a^\alpha=(0,\,0,\,0,\,-g(t))\qquad,\qquad k_\alpha=(0,\,0,\,0,\,-3\,f(t))\qquad.\label{Kfg}\end{align}
   This assumption results from the requirement that the contorsion tensor is invariant with respect to the isometry (sub)group of the metric : $\mathbf R^3\rtimes SO(3)$.  { It extends the framework considered in Ref. \cite{minkevich1980generalised} by  taking into account parity breaking terms.}\\
  Using this ansatz, there remains only four algebraically distinct field equations. One of them is a consequence of the other three, as expected from the Bianchi identities Eq.[\ref{BianchiId}]. The three relevant equations (see Ref. \cite{PhysRevD.95.024013}) ${\mathcal S}^1_{{\, .\,} 01}=0$, ${\mathcal S}^1_{{\, .\,}23}=0$, $\mathcal E^0_0=0$ lead to  :
 { \footnotesize{
  \begin{align}&c_R\,g(t)+   \kappa_N\,\sigma _{0^+} \,g(t)\,  \Big(g(t)^2 -f(t)^2  -\ddot A(t)-2\, \dot A(t)^2+3\,f(t) \, \dot A(t)+\dot f(t)\Big )\nonumber\\ &
   +(c_R\,\sigma_{2^-}-\kappa_N\,\sigma _{2^+})\,g(t)\,\Big(\ddot A(t)+\dot A(t)^2-f(t)\,\dot A(t)-\dot f(t)\Big)\nonumber\\
   &+c_R\, \sigma _{0^-}  \Big (g(t)\big(\ft 32\ddot A(t)+3\, f(t)\, \dot A(t)-\dot f(t)-2\,f^2(t)\big)+\ft 3 2\,\dot A(t)\,\dot g(t)+\ft 12\,\ddot g(t) \Big) =0\label{FRWK101}\\
  &  c_R\,f(t)+c_R\,\sigma_{0^-}\,g(t)\,\Big (g(t)\big(2\,f(t)-3\,\dot A(t)\big) -\dot g(t)\Big)+\kappa_N\,\sigma_{2^+}\,g(t)\big(g(t)\,\dot A(t)+\dot g(t)\bigg)\nonumber\\
  &+\kappa_N\,\sigma_{0^+}\Bigg(\bigg ( f(t)\Big((g^2(t)-f^2(t))+\ft 12 \ddot A(t)-2\,\dot A^2(t)+3\,f(t)\,\dot A(t)\Big) +g(t)\,\dot g(t)\bigg)+\ft3 2\,\dot f(t)\,\dot A(t)-2\,\dot A(t)\,\ddot A(t)+\ft 12 \ddot f(t)-\ft 12 \overset{\,\dots} A(t)\Bigg)  \nonumber\\  &
     -c_R\,\sigma_{2^-}\,g(t)\big(g(t)\,\dot A(t)+\dot g(t)\big)=0\label{FRWK123}\\
&\Lambda+3\,c_F\big(g^2(t)-f^2(t)+2\, f(t)\,\dot A(t)\big)-3\,\kappa_N\,\dot A^2(t)
+c_F\,\sigma_{0^-}\Big(\ft 12(3\,g(t)\,\dot A(t)-\dot g(t))^2- 2\,\dot g^2(t)+6\,f(t)\,g^2(t)( f(t)-2\,\dot A(t))\Big)\nonumber\\
&-\ft32\kappa_N\,\ft{c_F}{c_R}\bigg(f^2(t)\Big(f^2(t)-4\,f(t)\,\dot A(t)+5\,\dot A^2-2\,g^2(t)\Big) +2\,f(t)\,\dot A(t)\Big(\ddot A(t)-\dot A^2(t)-\dot f(t)+2\,g^2(t)\Big)
\nonumber\\
 &\phantom{-\ft32\kappa_N\,\ft{c_F}{c_R}\Big(} -\Big(\ddot A(t)+2\,\dot A^2(t)-\dot f(t)-g^2(t)\Big)\Big(\ddot A(t)-\dot f(t)+g^2(t)\Big) \bigg)=0\label{FRWE00}
\end{align}}}
We emphasise  that the trace of the Einstein equations leads to a remarkably simple equation, independent of the $\sigma_{J^\pm}$ parameters :
\begin{align}\Lambda+\ft 32\,c_F(g^2(t)-f^2(t))+\ft 92\,c_F\,f(t)\,\dot A(t)+\ft 32\,c_F\,\dot f(t)-3\,\kappa\,\dot A^2(t)-\ft 32 \kappa \,\ddot A(t)=0\qquad .\label{FRWTr}\end{align}
  In Ref.  \cite{PhysRevD.95.024013} cosmological solutions of class {$I\! I$}  torsion gravity equations, build on de Sitter geometry 
  \begin{align}ds^2=-dt^2+e^{2\,\lambda\,t}(dx^2+dy^2+dz^2)\label{metdS}\end{align}
  were studied, under the assumption that the contorsion tensor was of the special form Eqs [\ref{Kfg}].
  It was demonstrated  that this ansatz implies that the functions $f(t)$ and $g(t)$ are constants. The same proof remains valid when   a non-zero bare cosmological constant is included. This was analysed in Ref.  \cite{Nikiforova_2018}.  which provides torsionful solutions. \\
For class {$I$} Lagrangian, the consequences differ. Let us  briefly summarise   how the field equations are solved  in this framework. The equation ${\mathcal S}^1_{{\, .\,} 23}=0$  provides (assuming that $\kappa_N\,\sigma_{0^+}\neq c_r\,\sigma_{2^-}$) the expression of $\dot f(t)$ in terms of $f(t)$ , $g(t)$ and the parameters $\Lambda$, $\sigma_{0^+}$ and $\sigma_{2^-}$. From the equation ${\mathcal S}^1_{{\, .\,} 01}=0$ we obtain the expression of $\dot g(t)$ in terms of the same variables. Inserting them in the trace of the Einstein equations (Eq. [\ref{FRWTr}]) we obtain $g^2(t)$ as a function of $f(t)$ and the various parameters. The time derivative of this last relation, once the previously obtained expressions of the time derivatives of $f(t)$ and $g(t)$ are inserted in, gives a second relation linking $f(t)$ and $g(t)$. Its compatibility with the first ones fixes the bare cosmological constant to be :
\begin{align}\Lambda=3\,c_R\Big(\frac {c_F}{2\,\kappa_N\,\sigma_{0^+}-c_R\,\sigma_{2^-}}+\lambda^2\Big)
\end{align}
and leads to :
\begin{align}g^2(t)=-\frac{c_R}{2\,\kappa_N\,\sigma_{0^+}-c_R\,\sigma_{2^-}}+\big(f(t)-\lambda\big)^2\qquad .\label{g2dSC1}
\end{align}
Finally an elementary integration gives :
\begin{align}f(t)=f_0\,e^{-\lambda\, t}+\lambda-\frac{c_R}{\lambda\,(2\,\kappa_N\,\sigma_{0^+}-c_R\,\sigma_{2^-})}\qquad .
\end{align}
  Its worthwhile to notice that if $ {c_R}/({2\,\kappa_N\,\sigma_{0^+}-c_R\,\sigma_{2^-}})> 0$ the positivity of the righthand side of Eq. (\ref{g2dSC1}) restrict the range of the the time variable $t$ by requiring that $f_0\,e^{-\lambda\, t}$ is outside the interval  bounded by : $ {(c_R/\lambda)}/({2\,\kappa_N\,\sigma_{0^+}-c_R\,\sigma_{2^-}})\pm \sqrt{{c_R}/({2\,\kappa_N\,\sigma_{0^+}-c_R\,\sigma_{2^-}})}$ .   

More general exact solutions are less obvious to build. We obtain one, in class {$I\! I$}  theory, by restricting   the Lagrangian only to the spin 2 massive degrees of freedom, i.e. by putting $\sigma_{0^-}=0$ (in addition to the constraints Eqs.[\ref{cl2}]). Denoting by $Y(t):=f(t)-\dot A(t)$, and assuming $g(t)$ non identically zero, we obtain 
from Eq.[\ref{FRWK123}] and Eq.[\ref{FRWTr}] :
 \begin{align}&  \dot Y(t)=\frac{c_F}{\kappa_N\,\sigma^{2+}}-f(t)\,Y(t)-Y^2(t)\quad ,\label{Yeq}\\
&\dot f(t)=\frac2{3\,c_R}\,\Lambda-\frac 1{\sigma_{2^+}}+\frac {c_F}{c_R}\,g^2(t)-\frac{\kappa_N}{c_F}\,Y^2(t)-3\,f(t)\,Y(t)-2\,f^2(t) \qquad .\label{feq}
\end{align}
The trace of the Einstein equations [\ref{FRWE00}] gives us the function $g(t)$ :
\begin{align}g^2(t)=\frac 1{3\,c_F}\Big(3\,c_R\big(f (t)+Y(t)\big)^2 +3\, c_F\,Y^2(t)-\Lambda \Big)\qquad .\label{Zeq}
\end{align}
Substituting it  in the sum of the two   equations [\ref{Yeq}] and [\ref{feq}], we obtain :
\begin{align}\ddot A(t)= -\frac{c_F}{\kappa_N\,\sigma_{2^+}}+\frac{\Lambda}{3\,c_R}-\dot A^2(t)\qquad .
\end{align}
According to the sign of $ {\Lambda}/(3\,c_R)-c_F/(\kappa_N\,\sigma_{2^+})$     different solutions emerge (two integration constants being fixed by an appropriate coordinate choice) :
\begin{subequations}
\begin{align}& \text{Case \bf a :} \  \,\frac{\Lambda}{3\,c_R}-\frac{c_F}{\kappa_N\,\sigma_{2^+}}=:-\alpha^2<0   \qquad,\nonumber\\
 &\phantom{ \text{Case \bf a :} \  \,}A(t)=\ln[\cos(\alpha\, t)] \qquad,\qquad  f(t)=\frac 1{\alpha\,\cos(\alpha\,t)}\Big(f_0-\sin(\alpha\,t)\big(\alpha^2+\frac {c_R}{\kappa_N\, \sigma_{2^+}}\big)\Big)\ \,. \label{FLRWcasea}\\
 &\text{Case \bf b :}\ \, \frac{\Lambda}{3\,c_R}-\frac{c_F}{\kappa_N\,\sigma_{2^+}}=:\bar\alpha^2>0 \qquad,\nonumber\\ 
 &\phantom{ \text{Case \bf  b :} \  \,}A(t)=\ln[\cosh(\bar\alpha \,t)]\qquad,\qquad   f(t)=\frac 1{\bar\alpha\,\cosh(\bar\alpha\,t)}\Big(f_0+\sinh(\bar \alpha\,t)\big(\bar\alpha^2-\frac {c_R}{\kappa_N\, \sigma_{2^+}}\big)\Big)\ \, .\label{FLRWcaseb}
\end{align}
\end{subequations}
These metrics look like those of a spatially curved anti de Sitter or de Sitter spaces, excepted that here we have Euclidean flat space sections instead of hyperbolic planes.
 When $\Lambda\leq 0$, $g^2(t)\geq 0$ and  there is no  restriction on the domain of $t$ resulting from Eq. [\ref{Zeq}].\\ 
 In case of solutions of the type ({\bf a}) (Eq.[\ref{FLRWcasea}]) if   $3\,c_F\,c_R/(\kappa_N\,\sigma_{2^+})>\Lambda>0$ and $f_0^2> \Lambda(3\,c_R-\sigma_{2^+}\,\Lambda)/{(9\,c_R\,c_F\,\sigma_{2^+})}$  then $  t\in ]-\pi/(2\,\alpha),\, +\pi/(2\,\alpha)[$. But on the contrary to what occurs for anti de Sitter space the boundaries $t=\pm \pi/(2\,\alpha)$ constitute cosmological curvature singularity surfaces.\\
{ In case of solutions of type ({\bf b}) the function $g^2(t)$ is given by a ratio of two quadratic polynomials in the variable $\sinh(\alpha\,t)$. To discuss it  let us express it using the three parameters~:
\begin{align}
&\frac{c_F}{c_R}=:q>0\qquad,\qquad \bar \alpha^2\,\sigma_{2^+}=:\zeta>0\qquad,\qquad f_0\,\sigma_{2^+}=:\nu\qquad ; 
\end{align}
we obtain
\begin{align}
&g^2(t)=\frac{q\big(1-(1+q)\,\zeta)\big)\sinh^2(\alpha\,t)-2\,q\,(1+q)\,\nu\,\sinh(\alpha\,t)+(1+q)\big(q\,(1+q)\,\nu^2-\zeta\,(q+(1+q)\zeta)\big)}{q\,(1+q)^2\,\zeta\,\sigma_{2^+}\,\cosh^2(\alpha\,t)}\qquad .
\end{align}
The denominator is always positive. \\
If $\zeta<1/(q+1)$ the function $g^2(t)$ is positive near $t=\pm \infty$ but vanishes at $t=t_\pm$  where~:
\begin{align}
\sinh(\alpha\,t_\pm) = \frac{q\,(q+1)\,\nu\pm\sqrt{q \,(q+1)\,\zeta\big(q\,(q+1)^2\,\nu^2-((q+1)\,\zeta-1)((q+1)\,\zeta+q)\big)}}{q\,(1-\zeta\,(q+1))}\qquad .\label{tpm}
\end{align}
These two values $t_\pm$ of the $t$ coordinate define an interval where the contorsion is not defined, $g(t)$ being imaginary. Accordingly the solution cannot be considered for these values of the time coordinate, even if the metric remains without singularity. \\
 On the contrary if $\zeta>1/(q+1)$, the function $g(t)$ is not defined at $t=\pm \infty$. But if~:
 \begin{align}
 &\nu^2>\frac{((q+1)\,\zeta-1)((q+1)\zeta+q)}{q(q+1)^2}
 \end{align}
 the values of $t_\pm$ obtained from Eq. [\ref{tpm}] define a closed interval of time on which $g(t)$ is well defined (being real).
 
Let us emphasise that independently of the restriction on the contorsion the metric is well behaved.
 The space is everywhere regular. Its geometry interpolates between two asymptotic de Sitter geometries. The curvature remains bounded and the chart $\{x,y,z,t\}\in\mathbb R^4$ cover the all manifold  in the sense that it is geodesically complete.}
\subsubsection{A black hole solution}
Spherically symmetric and black holes solutions in torsion gravity are discussed intensively in Refs  \cite{damour2019spherically,nikiforova2020black}. They rest on a static spherically symmetric geometry written in Schwarzschild coordinates~:
\begin{align}ds^2=-e^{2\,A(r)}dt^2+e^{2\,B(r)}dr^2+r^2\,d\theta^2+r^2\,\sin(\theta)^2\,d\varphi^2
\end{align}
and a contorsion tensor invariant to the complete isometry group of the metric, including time-reversal and parity. This restricts the non-zero contorsion components to two independent ones, parametrised as follows : 
\begin{align} &K_{rtt}=-K_{trt}=e^{2\,A(r)-B(r)} \big(v(r)-e^{-B(r)}\big)
\end{align}
and
\begin{align}& K_{r\theta\theta}= \csc^2(\theta)\,K_{r\varphi\varphi}=K_{\theta r\theta}=-\csc^2(\theta)\,K_{\varphi r \varphi}=\big(r^2\,e^{B(r)}\,w(r)+  r\big)\qquad.
\end{align}  
 We will assume that in addition to the previous components there is also a non-time reversal invariant term :
\begin{align}K_{trr}=e^{A(r)+2\,B(r)}\,P(r)\qquad.
\end{align}
Such a term is similar to the electric field of the Reissner-Nordstrom solution.\\
Again the two possible classes of models lead to completely different solutions. Let us first consider class $I$ models, those where $\sigma_{0^-}$ and $\sigma_{2^+}$ are set equal to zero.\\
Equation ${{\mathcal S}}^{rtr}=0$ reads :
\begin{align} e^{-A(r)-2\,B(r)}\big({c_F\,\sigma_{2^-}-  \frac{2\,c_F}{3\,c_R}\,\kappa_N\sigma_{0^+}-2\,\phi\big)\,P(r)\,w^2(r)=0\qquad .}\label{SrtrCI}
\end{align}
Among the two solutions of this equation   :
\begin{align}w(r)=0
\end{align}
is the simpler. It constitutes also a solution of the Einstein equation $\mathcal E^t_r=0$. Inserted in the other equations, we obtain from the difference $\mathcal E^t_t-\mathcal E^r_r=0$ :
\begin{align}A(r)=-B(r)+A_0\qquad .\label{ABcI}
\end{align}
The integration constant $A_0$ can, as usual, be eliminated by a rescaling of the time coordinate. Then, from the connection equation $\mathcal S^\theta_{{\, .\,} t\theta}=0$ and the Einstein equation $\mathcal E^r_r=0$ we obtain the first order differential system :
\begin{align}&v'(r)=\frac 1{4\,c_R\,\kappa_N\,\sigma_{0^+}}\Bigg(4\,c_R\bigg( \frac{\kappa_N\,\sigma_{0^+}}{r^2}-3\,c_R\bigg)e^{B[r]}\nonumber\\
&\phantom{v'(r)= }+\frac 1r\bigg(2\,c_R\,\kappa_N\,\sigma_{0^+}-\Big(\big(c_R\,(2\,\kappa_N\,\sigma_{0^+}+3\,c_F\,r^2)-2\,\kappa_N\,\sigma_{0^+}\,\Lambda\,r^2\big)e^{2\,B(r)}\Big)\bigg)v(r)\Bigg)\qquad ,\\
&B'(r)=\Bigg(\bigg(\frac{\Lambda}{2\,c_R}-\frac{3\,C_F}{4\,\kappa_N\,\sigma_{0^+}}\bigg)r-\frac 1{2\,r}\Bigg)e^{2\,B(r)}+\frac 1{2\,r}\qquad ,
\end{align}
whose solution is :
\begin{align}&e^{-2\,B(r)}=1-\frac {2\,\mu}r+\Big(\frac{c_F}{2\,\kappa_N\,\sigma_{0^+}}-\frac{\Lambda}{3\,c_R}\Big)\,r^2\qquad ,\label{BcI}\\
&v(r)=\Big(v_0-\frac{3\,c_R}{\kappa_N\,\sigma_{0^+}}\,r+\frac 1r\Big)\, e^{B(r)}\qquad .\label{vcI}
\end{align}
The functions $A(r)$, $B(r)$, $v(r)$ given by Eqs [\ref{ABcI}, \ref{BcI}, \ref{vcI}] with $w(r)=0$ and $P(r)$ that remain arbitrary constitute a solution of the complete system of Einstein and connection field equations. Note that the contorsion component $v(r)$ only depends on   one arbitrary constant $v_0$ that allows to make it regular on the black hole horizon but, in general, not also on a cosmological horizon. \\
In case we fix the coupling constant 
\begin{align}
\phi=\frac 12\,c_F\,\sigma_{2^-}- \frac{c_F}{3\,c_R}\,\kappa_N\,\sigma_{0^+}
\end{align}
in order to solve Eq. [\ref{SrtrCI}], the Lagrangian on shell becomes independent of $\sigma_{2^+}$. The functions $P(r)$ and $w(r)$ remain arbitrary. The metric components are given by
 \begin{align}
e^{2\,A(r)}=e^{-2\,B(r)}=1-2\,\frac{\mu}r+\Big(\frac{c_F}{2\,\kappa_N\,\sigma_{0^+}}-\frac{\Lambda}{3\,c_R}\Big)r^2\qquad ,
\end{align}
$\mu$ being an integration constant, while another is fixed by a rescaling of the $t$ coordinate. The last unknown function $v(r)$ has to be the solution of the first order differential equation 
\begin{align}
&v'(r)=\bigg(2\,e^{B(r)}+B'(r)\bigg)v(r)+\bigg(\frac 1{r^2}-3\,\frac{c_R}{\kappa_N\,\sigma_{0^+}}\bigg)\,e^{B(r)}\nonumber\\
&\phantom{v'(r)=}+\bigg(\frac {2\,w(r)}r+2\,w'(r)-e^{B(r)}\,w^2(r)\bigg)\qquad .
\end{align}

In the framework of models of class $I\! I$, the coupling constant $\sigma_{0^-}$ do not appear in the field equations. We  first obtain
\begin{align} 
&e^{-A(r)-2\,B(r)}\big(2\,\phi+  \frac{ c_F}{3\, c_R}\,\kappa_N\sigma_{2^+}\big)\,P(r)\,w^2(r)=0\qquad .\label{StrtCII}
\end{align}
A strategy similar as the previous leads also to Eq. [\ref{ABcI}] and
\begin{align}&e^{-2\,B(r)}=\frac{c_R+3\,c_F}{c_R}-2\,\frac{\mu}r-\frac{(6\,c_F\,c_R+\kappa_N\,\sigma_{2^+}\,\Lambda)}{3\,\kappa_N\,c_R\,\sigma_{2^+}}\,r^2\qquad ,\\
&v(r)=\Big(v_0-\big(\frac{6\,c_R}{\kappa_N\,\sigma_{2^+}}r+\frac 1r\big)\Big)\,e^{B(r)} 
\end{align}
but from the equation ${\mathcal S}^{rtt}=0$ we obtain :
\begin{align}c_F\,\frac {e^{A_0}}r=0\qquad .
\end{align}
Thus, for consistency, we have to put the coupling constant $c_F$ equal to zero,  which constitutes a physically unacceptable condition. The same conclusion occurs if instead of fixing $w(r)=0$ to solve Eq. [\ref{StrtCII}] we fix the coupling constant $\phi=-c_F/(6\,c_R)\,\kappa_N\,\sigma_{2^+}$.\\
   {A lot of endeavour have been devoted to the study of Birkhoff's theorem in the framework of quadratic torsion gravity (see for instance Refs \cite{ramaswamy1979birkhoff,rauch1982birkhoff} and especially \cite{Obukhov:2020hlp} and references therein). To summarise the result of this section, for class $I$ models we obtain a spherically symmetric configuration which is not the torsionless Schwarzschild solution. The geometry is the Schwarzschild-de Sitter metric but the torsion involves arbitrary functions. Accordingly Birkhoff's theorem is not satisfied in this framework. However for class $I\! I$ theories everything seems to fall into place (see Ref.\cite{nikiforova2020black}). To make an end we also want to mention Ref. \cite{baekler1988hamiltonian} devoted to the Hamiltonian approach of the theory. This works obtains several very interesting examples of torsionful  spherically symmetric solutions, but  the one presented here above seems to have escaped.}

To conclude let us mention that on-shell the Lagrangian reduces to its Einstein part, the contorsion term $L_F$ vanishes and the effective cosmological constant appearing in the metric reduces to its bare value. Thus there is no one-loop quantum contribution from the contorsion expected for this black hole configuration.
\section{Conclusion}
We studied pure torsion gravity theories without matter sources. This simplification has allowed us to write the field equations  in terms
 of the metric and contorsion components expressed with respect to a natural (coordinate) frame instead of vielbiens and spin coefficients.     Of course both approaches are equivalent, but the former is simpler than the latter (and well adapted for symbolic calculations on the computer). We also wrote a general expression of the Noether identities resulting from the diffeomorphism invariance of the theory. These identities, established, in a general framework, look more useful in this context than the Bianchi ones.\\
 To obtain specific solution of the theory we restricted ourself   to quadratic models. Two classes of physically acceptable such models are known.   They mainly differ by the parity of their respective spin zero and spin two massive fields.  We  have obtained analytical solutions of the field equations in various contexts.     Some of them present an unexpected aspect that make the  theory questionable. The contorsion field (which is not a gauge field) sometime involves arbitrary functions, that may even be time dependent, see Eq. [\ref{Atz}] but dont play any r\^ole in the expression of the spacetime metric. 
{The occurrence of arbitrary functions in the expression of some of the solutions presented here above raises the question of the predictability of the theory and thus the possibility of a possible confrontation of the models with observations. This point has already been raised by various authors. For instance, predictability is discussed in reference \cite{dimakis1989initial} where sufficient conditions of uniqueness of the solution of the Cauchy problem are established\footnote{As noticed in Ref. \cite{hecht1996some}, models of class $I$ and $I\! I$  do not verify them}, but some  Lagrangians with well-posed initial value problems, that escape these sufficient conditions are described in Ref. \cite{hecht1996some}. 
 Moreover, various authors   have also obtained solutions containing arbitrary functions : in the framework of an analysis of asymptotic solutions \cite{chen1994poincare} or in the ones of solutions built using the  double duality  ansatz \cite{ lenzen1985spherically, zhytnikov1996double}. In particular, this last work proposes an interesting conjecture. The appearance of arbitrary functions would reflect a hidden gauge symmetry that could be revealed by a Hamiltonian analysis, as the symmetry could emerge through a bifurcation phenomenon of the constraint algebra for certain configurations. However, none of the arbitrary functions  encountered (that are not reflecting the diffeomorphism invariance) are unrestricted. They do not depend on all the coordinates, which makes the previous interpretation unlikely. In any case, the question of their physical and mathematical meaning remains open and requires a further (certainly difficult) work to be elucidated.}
 
 We also met obstruction to the existence of the contorsion field despite the fact that the metric remains perfectly regular (see the solution Eq. [\ref{FLRWcaseb}], in section [{\bf\ref{FLRW}}]). The specific solutions we obtain also emphasise differences between the two classes of quadratic gravity theories. For instance we obtain, under specific assumption,  a black hole solution for the class $I$ theory that cannot exits in the framework of the class $I\! I$.
  \appendix
\section{Euler invariant and Bach-Lanczos identity\label{GBE}}
In the main text we made use of a topological invariance of the Euler  class to obtain Eq.[\ref{Eclass}] and a quadratic identity satisfied by the Riemann curvature tensor to pass from Eq.[\ref{Del1}]) to Eq. [\ref{Del2}]. In this appendix, for the reader convenience, we   sketch a proof of these properties.\\
To start we notice that the Pfaffian   {(exterior products of connection one-forms $\underline A^\alpha_{{\, .\,}\beta}=A^\alpha_{{\, .\,}\beta\mu}\,dx^\mu$, and curvature two-forms $\underline{\underline{F}}^{\alpha\beta}=\ft 12 \, {F}^{\alpha\beta}_{{\, .\,}{\, .\,}\mu\nu}dx^\mu\wedge dx^\nu$ are implied) :}
\begin{align}\mathbf{\Omega_4}=\ft 12\,\eta_{\alpha\beta\gamma\delta}\,\underline{\underline{F}}^{\alpha\beta}\underline{\underline{F}}^{\gamma\delta}\,=\ft 12\, \epsilon^{\hat a\hat b\hat c\hat d} \,\underline{\underline{F}}^{\hat a\hat b}\underline{\underline{F}}^{\hat c\hat d}
\end{align}
is an invariant polynomial, such that  $d\mathbf{\Omega_4}=0$ (as there is no 5-forms in four dimensions). Accordingly it can be written, locally, as an exact differential. Using the well-known variation trick (homotopy operator \cite{carterCargese1978,Zumino:1983ew}) we obtain~:
\begin{align}\mathbf{\Omega_4}=d\mathbf{\Omega_3}\label{W4dW3}
\end{align}
with
\begin{align}&\mathbf{\Omega_3}=\ft 12\epsilon^{\hat a\hat b\hat c\hat d}\underline A_{\hat a\hat b}{\big(}\underline{\underline F}_{\hat c\hat d}-\ft13\,\underline A_{\hat c\hat k} \,\underline A^{\hat k}_{{\, .\,} \hat d}\big)
 =\ft 12 \,\eta^{\alpha\beta\gamma\delta}\underline A_{\alpha\beta}{\big(}\underline{\underline F}_{\gamma\delta}-\ft13\,\underline A_\gamma^{{\, .\,} \rho}\,\underline A_{\rho\delta}\big)
\end{align}
For a direct check of Eq.(\ref{W4dW3}) we use the lemma :\\
{\it Lemma} : If $A^{\alpha_i\beta_j}_{\tau_k}=-A^{\beta_j\alpha_i }_{\tau_k}$ then, in  dimension $n$,
\begin{align}B^{\beta_1\cdots\beta_n}:=A^{\alpha_1\beta_1}_{\tau_1}\cdots A^{\alpha_n\beta_n}_{\tau_n}\epsilon^{\tau_1\cdots  \tau_n}\,\epsilon_{\alpha_1\cdots\alpha_n}=0\label{Lem}
\end{align}
which implies in particular in four dimensions that ($ {d^4x}$ denotes the affine-volume 4 form)
\begin{align}\epsilon_{\hat a\hat b\hat c\hat d}\underline A^{\hat a\hat k_1}\underline A^{\hat b\hat k_2}\underline A^{\hat c\hat k_3}\underline A^{\hat d\hat k_4} =\epsilon_{\hat a\hat b\hat c\hat d}  A^{\hat a\hat k_1}_\mu A^{\hat b\hat k_2}_\nu A^{\hat c\hat k_3}_\rho A^{\hat d\hat k_4}_\sigma\,\epsilon^{\mu\nu\rho\sigma} {d^4x}=0\qquad .
\end{align}
{\it Proof} : First let us notice the symmetry of $B^{\beta_1\cdots\beta_n}=B^{(\beta_1\cdots\beta_n)}$. But on the other hand writing $A^{\alpha_1\beta_1}_{\tau_1}$ as $\epsilon^{\alpha_1\beta_1\kappa_3\cdots\kappa_n}T_{\kappa_3\cdots\kappa_n,\tau_1}$ we obtain :
\begin{align}{&B^{\beta_1\cdots\beta_n}=\epsilon^{\tau_1\cdots\tau_n}\epsilon_{\alpha_1\cdots\alpha_n}\epsilon^{\alpha_1\beta_1\kappa_3\cdots\kappa_n}T_{\kappa_3\cdots\kappa_n,\tau_1}A^{\alpha_2\beta_2}_{\tau_2}\cdots A^{\alpha_n\beta_n}_{\tau_n}\\
&\phantom{B^{\beta_1\cdots\beta_n}}=\epsilon^{\tau_1\cdots\tau_n}\delta^{\beta_1}_{[\alpha_2}\delta_{\alpha_3}^{\kappa_3} \cdots\delta_{\alpha_n]}^{\kappa_n}T_{\kappa_3\cdots\kappa_n,\tau_1}A^{\alpha_2\beta_2}_{\tau_2}\cdots A^{\alpha_n\beta_n}_{\tau_n}\\
&\phantom{B^{\beta_1\cdots\beta_n}}=\epsilon^{\tau_1\cdots\tau_n}T_{\alpha_2\cdots\hat \alpha_k\cdots\alpha_n,\tau_1}{\big(}\sum_{k=2}^n (-)^k\,A^{\alpha_2\beta_2}_{\tau_2}
\cdots A^{\beta_1\beta_k}_{\tau_k}\cdots A^{\alpha_n\beta_n}_{\tau_n}\big) 
}\end{align}
But as  $B^{\beta_1\cdots\beta_n}$ is completely symmetric in the indices $ \beta_1\cdots\beta_n$ while $A^{\beta_1\beta_k}=-A^{\beta_k\beta_1}$ we deduce that $B^{\beta_1\cdots\beta_n}=0$.

Accordingly :
\begin{align}\mathbf{\Omega_4}=\ft 12\epsilon^{\hat a\hat b\hat c\hat d}(d\underline A_{\hat a\hat b}\,d\underline A_{\hat c\hat d}+2\,d\underline A_{\hat a\hat b}\,\underline A_{\hat c\hat k}\underline A^{\hat k}_{{\, .\,} \hat d})\label{W4dA}
\end{align}
whereas
\begin{align}{&\ft 12\epsilon^{\hat a\hat b\hat c\hat d}d\Big(\underline A_{\hat a\hat b}{\big(}\underline{\underline F}_{\hat c\hat d}-\ft13\,\underline A_{\hat c\hat k} \,\underline A^{\hat k}_{{\, .\,} \hat d}\big)\Big)=\ft 12\epsilon^{\hat a\hat b\hat c\hat d}d\Big(\underline A_{\hat a\hat b}{\big(}d{\underline A}_{\hat c\hat d}+\ft 23\,\underline A_{\hat c\hat k} \,\underline A^{\hat k}_{{\, .\,} \hat d}\big)\Big)\\
&=
\ft 12\epsilon^{\hat a\hat b\hat c\hat d}\Big(d\underline A_{\hat a\hat b}\,d{\underline A}_{\hat c\hat d}+\ft 23\,d\underline A_{\hat a\hat b}\,\underline A_{\hat c\hat k} \,\underline A^{\hat k}_{{\, .\,} \hat d}-\ft 43 \,\underline A_{\hat a\hat b}\,d\underline A_{\hat c\hat k} \,\underline A^{\hat k}_{{\, .\,} \hat d}\Big)}\end{align}
  Schouten's lemma implies that :
  \begin{align}\epsilon^{\hat a\hat b\hat c\hat d}( \underline A_{[\hat a\hat b}\,d\underline A_{\hat c\hat k} \,\underline A^{\hat k}_{{\, .\,} \hat d]})=0=4\,\epsilon^{\hat a\hat b\hat c\hat d}(\underline A_{[\hat a\hat b}\,d\underline A_{\hat c\hat k]} \,\underline A^{\hat k}_{{\, .\,} \hat d})\qquad .
\end{align}
Accordingly
\begin{align}\epsilon^{\hat a\hat b\hat c\hat d}(\underline A_{\hat a\hat b}\,d\underline A_{\hat c\hat k} \,\underline A^{\hat k}_{{\, .\,} \hat d}-\underline A_{\hat b\hat c}\,d\underline A_{\hat k\hat a} \,\underline A^{\hat k}_{{\, .\,} \hat d}+\underline A_{\hat c\hat k}\,d\underline A_{\hat a\hat b} \,\underline A^{\hat k}_{{\, .\,} \hat d}+\underline A_{\hat k\hat a}\,d\underline A_{\hat b\hat c} \,\underline A^{\hat k}_{{\, .\,} \hat d})=0
\end{align}
 i.e.  
\begin{align}&\epsilon^{\hat a\hat b\hat c\hat d}\,\underline A_{\hat a\hat b}\,d\underline A_{\hat c\hat k} \,\underline A^{\hat k}_{{\, .\,} \hat d}=-\epsilon^{\hat a\hat b\hat c\hat d}\,\underline A_{\hat c\hat k}\,d\underline A_{\hat a\hat b} \,\underline A^{\hat k}_{{\, .\,} \hat d} =-\epsilon^{\hat a\hat b\hat c\hat d}\,d\underline A_{\hat a\hat b}\,\underline A_{\hat c\hat k} \,\underline A^{\hat k}_{{\, .\,} \hat d} \qquad\qquad \qed
\end{align}

The topological invariant obtained by integration of $\mathbf{\Omega_4}$ reads :
\begin{align}\int \mathbf{\Omega_4}=\ft 12\int  F^{\alpha\beta}_{{\, .\,}{\, .\,} \rho\sigma}\,F^{\gamma\delta}_{{\, .\,}{\, .\,} \omega\tau}\,\eta_{\alpha\beta\gamma\delta}
\,\eta^{\rho\sigma\omega\tau}\sqrt{-g}\,d^4x=\ft 1{12}\int {\big(}F^{\alpha\beta}_{{\, .\,}{\, .\,} \gamma\delta}\,F^{\gamma\delta}_{{\, .\,}{\, .\,} \alpha\beta}-4\,F^{\alpha}_{{\, .\,} \beta}\,F^{\beta}_{{\, .\,} \alpha}+F^2\big)\sqrt{-g}\,d^4x\quad .
\end{align}
The Bach-Lanczos identity \cite{Bach:1921aa,Lanczos1938}, that is used in the main text, can obtained by computing, assuming the  connection torsionless (Levi-Civita connection), the variation of the $\int \mathbf{\Omega_4}$ with respect to the metric. Using $\delta R^{\alpha}_{{\, .\,} \beta\mu\nu}={\nabla\!}_\mu\delta \Gamma^\alpha_{{\, .\,}\beta\nu}-{\nabla\!}_\nu\delta \Gamma^\alpha_{{\, .\,}\beta\mu}$ and the Bianchi identity ${\nabla\!}_\rho R^{\alpha}_{{\, .\,} \beta\mu\nu}\,\eta^{\mu\nu\rho\sigma}\equiv 0$ we deduce that~:
\begin{align}R_{\alpha\mu\nu\rho}\,R_{\beta}^{{\, .\,} \mu\nu\rho}=2\,R_{\alpha\mu\beta\nu}\,R^{\mu\nu}+2\,R_{\alpha\mu}R^\mu_\beta-R\,R_{\alpha\beta}+\ft 14 \,g_{\alpha\beta}(R_{\mu\nu\rho\sigma} R^{\mu\nu\rho\sigma}-4\,R_{\mu\nu}R^{\mu\nu }+R^2)
\end{align}
This identity generalises straightforwardly to $2\,n$ dimensions. Restricting ourselves to Riemann-curvature ${\mathbf{\Omega}}_4$ generalises to
\begin{align}{\mathbf{\Omega}}_{2n}=\frac{\sqrt{-g}}{n}\delta^{[\mu_1}_{\alpha_1}\delta^{\nu_1\cdots}_{\beta_1\cdots} \delta^{\mu_n}_{\alpha_n}\delta^{\nu_n]}_{\beta_n}\,R^{\alpha_n\beta_n}_{\,{\, .\,}\, {\, .\,} \,\mu_n\nu_n}\cdots R^{\alpha_1\beta_1}_{\,{\, .\,} \,{\, .\,}\, \mu_1\nu_1}
\end{align}
which is a divergence in $2\,n$ dimensions and whose Euler-Lagrange  variation leads to the identity :
\begin{align}\delta^{[\mu_1}_{\alpha_1}\delta^{\nu_1\cdots}_{\beta_1\cdots} \delta^{\mu_n}_{\alpha_n}\delta^{\nu_n]}_{(\rho}\,R^{\alpha_n}_{\,{\, .\,} \,\sigma)\mu_n\nu_n}\cdots R^{\alpha_1\beta_1}_{\, {\, .\,}\,{\, .\,} \,\mu_1\nu_1}=\frac 1{2\,n}\,g_{\rho\sigma}\,\delta^{[\mu_1}_{\alpha_1}\delta^{\nu_1\cdots}_{\beta_1\cdots} \delta^{\mu_n}_{\alpha_n}\delta^{\nu_n]}_{\beta_n}\,R^{\alpha_n\beta_n}_{\,{\, .\,}\, {\, .\,} \,\mu_n\nu_n}\cdots R^{\alpha_1\beta_1}_{\,{\, .\,} \,{\, .\,}\, \mu_1\nu_1}\qquad .
\end{align}
For $n=1$ it reduces to the well known relation between the Ricci tensor and the scalar curvature~: $R_{\alpha}^\beta=\ft 12 \,\delta_{\alpha}^\beta\,R$. In dimensions different from $2\,n$ all of these ``invariants" constitute the building blocks of the Lovelock gravity theory  \cite{lovelock1971einstein} .

\section{Conventions\label{conv}}
  {For the readers convenience we have emphasised the density nature of some objects by underlying their symbol with a dot. 
Our conventions for the contractions of the Riemann and of the curvature   tensors are~:
\begin{align}
&R_{\alpha\beta}:= R^\mu_{{\, .\,} \alpha\mu\beta}=-R^{{\, .\,}\mu}_{\alpha{\, .\,} \mu\beta}=-R^\mu_{{\, .\,} \alpha\beta\mu}= R_{ \beta \alpha}
\qquad,\\
& R:= g^{\alpha\beta}R_{\alpha\beta}\qquad ,\\
&F_{\alpha\beta}:= F^\mu_{{\, .\,} \alpha\mu\beta}=-F^{{\, .\,}\mu}_{\alpha{\, .\,} \mu\beta}=-F^\mu_{{\, .\,} \alpha\beta\mu}\neq F_{ \beta \alpha}
\end{align}
as, on the contrary to the Riemann tensor who verifies the relation : $R_{\alpha\beta\gamma\delta}= R_{\gamma\delta\alpha\beta} $, in general $g_{\alpha\mu}F^\mu_{{\, .\,}\beta\gamma\delta}:=F_{\alpha\beta\gamma\delta}\neq F_{\gamma\delta\alpha\beta} $.}\newline
We summarise in the next table the relationships between ours notations and those used by some authors to denote the coupling constants used in the quadratic Lagrangian they consider. We indicate  some of the restrictions they impose on their parameters in the caption of the table. To establish these correspondences we have made use the Euler  class discussed in the previous Appendix ( $\sim$ meaning an equality up to a divergence) :
\begin{align}F^2&\sim  4\,F_{\alpha\beta}F^{\beta\alpha}-F_{\alpha\beta\gamma\delta}F^{\gamma\delta\alpha\beta}\qquad ,\label{Eclass}
\end{align}
and the expression of the scalar Riemann-curvature obtained from a double contraction of Eq. [\ref{F4R4}] followed by the substitution in it of the contorsion in terms of the torsion (Eq. [\ref{TK}]) :
 \begin{align} F&=R+2\,{\nabla\!}_\rho K^{\rho\sigma}_{{\, .\,}\,{\, .\,}\,\sigma}+K^\rho_{{\, .\,} \sigma\rho}K^{\sigma\tau}_{{\, .\,}\,{\, .\,}\,\tau}-K^\rho_{{\, .\,} \sigma\tau}K^{\tau\sigma}_{{\, .\,}\,{\, .\,}\,\rho}
\end{align}
 Thus we obtain   :
 \begin{align}&F\sim  R+\ft 14 \,T_{\alpha\beta\gamma}T^{\alpha\beta\gamma}+\ft 12\,T_{\alpha\beta\gamma}T^{\gamma\beta\alpha}-T_\alpha\,T^\alpha\qquad 
\end{align}  
and 
 \begin{align}&{c_R}\,R+{c_F}\,F\sim {\kappa_N}F-{c_R}(\ft 14 \,T_{\alpha\beta\gamma}T^{\alpha\beta\gamma}+\ft 12\,T_{\alpha\beta\gamma}T^{\gamma\beta\alpha}-T_\alpha\,T^\alpha)\qquad .
\end{align}
Using the torsion decomposition into irreducible parts :
 \begin{align}&T_\alpha=T^\beta_{{\, .\,} \beta\alpha}\label{trT}\\
 &t_{\alpha\beta\gamma}=\ft12(T_{\alpha\beta\gamma}+T_{ \beta\alpha\gamma})+\ft 16(g_{\alpha\gamma}T_\beta+g_{\beta\gamma}T_\alpha)-\ft 13 g_{\alpha\beta}T_\gamma\label{T16}\\
& a^\delta=\ft 16 \eta^{\delta\alpha\beta\gamma}T_{\alpha\beta\gamma}\label{T4}
\end{align}
we obtain
 \begin{align}&t_{\alpha\beta\gamma}t^{\alpha\beta\gamma}=\ft12(T_{\alpha\beta\gamma}T^{\alpha\beta\gamma}+T_{\alpha\beta\gamma}T^{ \gamma\alpha\beta}) -\ft 12 T_\alpha T^\alpha\\
 &a^\delta a_\delta=\ft 1{18}(2\,T_{\alpha\beta\gamma}T^{\gamma\beta\alpha}- T_{\alpha\beta\gamma}T^{\alpha\beta\gamma}) \end{align}
{i.e. }
\begin{align}&{c_R}\,R+{c_F}\,F\sim  {\kappa_N}F-{c_R}( \ft 23\, t_{\alpha\beta\gamma}t^{\alpha\beta\gamma}+\ft 32\, a_\alpha a^\alpha-\ft 23\, T_\alpha T^\alpha)\qquad .
\end{align}
In other words we fix the coupling constants of the terms quadratic in the torsion field instead of  leaving them arbitrary as its is the case in Refs   \cite{PhysRevD.21.3269,PhysRevD.24.1677,Hayashi1980gravity1,Hayashi1980gravity2,Hayashi1980gravity3,Hayashi1980gravity4}.\\
Let us mention that some authors  use  the square of $\star F:=\ft 1{4!}\eta^{\alpha\beta\mu\nu}F_{\alpha\beta\mu\nu}$ in the expression of the Lagrangians they consider. To make contact with ours, we remind that~:
 \begin{align} (\star F)^2&= -\ft 4{( \,4!\,)^2}(F_{\alpha\beta\gamma\delta}F^{\alpha\beta\gamma\delta}-4\,F_{\alpha\beta\gamma\delta}  F^{ \alpha\gamma\beta\delta}+F_{\alpha\beta\gamma\delta}F^{\gamma\delta\alpha\beta})\qquad .\label{stF2}\end{align}
 
 \newpage

 {\strut  {\begin{table}[tbp]{\footnotesize{\begin{tabular}{|c|c|c|c|c|c|c|}
\hline
& {\it I}& {\it II}& {\it III}\footnotemark
&{\it IV}&{\it V}&{\it VI}\rule[-7pt]{0pt}{20pt} \\ \hline 
$F$&$ 
  {c_R+c_F=:\kappa_N}$
&
$\lambda$&$ a$&$c_1$&$  {\ft32(\tilde\alpha+\overline\alpha)=\kappa_N}$\rule[-7pt]{0pt}{20pt}&$  {c_R+c_F=\kappa_N}$\\
\hline
$t_{\alpha\beta\gamma}t^{\alpha\beta\gamma}$&$-\ft 23{c_R}$&$-\ft 23(\lambda+ a)$&$\alpha $&$\alpha $&$-\overline\alpha  $\rule[-7pt]{0pt}{20pt}&$-\ft 23\,c_R $\\
\hline
$a_\alpha a^{\alpha}$&$-\ft 32{c_R}$&$-\ft 32\, (\lambda-b) $&$\gamma $&$\gamma $&$ -\ft 94\overline\alpha $\rule[-7pt]{0pt}{20pt}&$-\ft 32\,c_R  $\\
\hline
$T_{\alpha }T^{\alpha }$&$+\ft 23{c_R}$&$\ft 23(\lambda-c) $&$\beta $&$\beta $&$ \overline\alpha $\rule[-7pt]{0pt}{20pt}&$ +\ft 23\,c_R$\\
\hline
$F_{\alpha\beta}F^{\alpha\beta}$ & $\ft 12 f_1$ & $(s+t)$ & $b_3$&$  c_3$\rule[-7pt]{0pt}{20pt}& $  c_3 $& $\ft 12(c_{F^2}+c_{34})$\\
\hline
$F_{\alpha\beta}F^{\beta\alpha}$ & $\ft 12 f_2$ & $(s-t)$ & $b_4+ 4\,b_5$&$   c_4+4\,  c_5$\rule[-7pt]{0pt}{20pt}&$   c_4+4\,  c_5$&  $-(\ft12\,c_{34}+\ft 56\,c_{F^2})$\\ 
\hline
$ F_{\alpha\beta\gamma\delta}F^{\alpha\beta\gamma\delta}$ & $\ft 16 d_1$ & $\ft 16(2\,p+q)$ & $b_1-4\,b_6$&$  b-4\,  c_6$ \rule[-7pt]{0pt}{20pt}&$   -4\,  c_6$& 0 \\
\hline
$F_{\alpha\beta\gamma\delta}F^{\alpha\gamma\beta\delta}$ & $\ft 16 d_2$ & $\ft 23(p-q)$ & $16\,b_6$&$16\,  c_6$\rule[-7pt]{0pt}{20pt}  &$16\,  c_6$&  0\\
\hline
$F_{\alpha\beta\gamma\delta}F^{ \gamma \delta\alpha\beta}$ & $\ft 16 d_3$ & $\ft 16(2\,p+q-6\,r)$ & $b_2-b_5-4\,b_6$&$-(  c_5+4\,  c_6)$\rule[-7pt]{0pt}{20pt}&$-(  c_5+4\,  c_6)$ &$\ft 13\,c_{F^2}$ \\ 
\hline
\end{tabular}}\vspace{1mm}\\
{ \hspace{-9mm}$\strut ^a$ The parameter $b_5$ is redundant. It can be eliminated by the redefinitions $b_4\mapsto b_4-4\,b_5$, $b_2\mapsto b_2+b_5$.} 
\caption{{ \it I} : this text (  {$\kappa_N:=1/16\,\pi\,G$}); {\it II} : Sezgin \cite{PhysRevD.24.1677} ; {\it III} : Hayashi \& Shirafuji \cite{Hayashi1980gravity4} ; {\it IV} : Nair \& al.\cite{PhysRevD.80.104031} ($\beta=-\alpha$, $\gamma=\ft 94 \alpha$, $b=0$, $c_5=-\ft 13(c_3+c_4)$;) , Nikiforova {\it et al}. \cite{PhysRevD.95.024013} ($c_5=-\ft 13(c_3+c_4)$, $c_5+16\,c_6<0$) ;  {\it V}~: Damour \& Nikiforova \cite{Nikiforova_2018} ($c_5=-\ft 13(c_3+c_4)$) ; {\it VI} :  Damour \& Nikiforova \cite{damour2019spherically},}}
\end{table}
}}
Models {\it IV}, {\it V} and {\it VI} are all of  class $I\! I$.
\newpage
 \section*{Acknowledgements}
I'm grateful to N. Boulanger, T. Damour and S. Massar for enlightening useful discussions and suggestions. This work is partially supported by the IISN convention 4.4503.15 .


\begin{thebibliography}{69}
\bibitem{baekler1986nonmetricity}
P.~Baekler, F.~W. Hehl, and E.~W. Mielke, {  `` Nonmetricity and torsion: Facts
  and fancies in gauge approaches to gravity"}.
\newblock R. Ruffini (ed.), {\it Proc. of the 4 th {M}arcel {G}rossmann Meting
  on {G}eneral {R}elativity}, Elsevier Science Publ., Amsterdam, 1986.

\bibitem{blagojevic2013gauge}
M.~Blagojevic and F.~Hehl, ``{G}auge theories of gravitation : a {R}eader
  with {C}ommentaries", reprint edn, M. Blagojevi\'c, F. Hehl (eds), Imperial College Press, London , 2013.

\bibitem{Clifton_2012}
T.~Clifton, P.~G. Ferreira, A.~Padilla, and C.~Skordis, ``Modified gravity and
  cosmology,'' {  \it Physics Reports}, vol.~513, pp.~1--189, Mar 2012.

\bibitem{de_Rham_2014}
C.~de~Rham, ``Massive gravity,'' {  \it Living Reviews in Relativity}, vol.~17, 7,
  Aug 2014.

\bibitem{PhysRevD.21.3269}
E.~Sezgin and P.~van Nieuwenhuizen, ``New ghost-free gravity Lagrangians with
  propagating torsion,'' {  \it Phys. Rev. D}, vol.~21, pp.~3269--3280, Jun 1980.

\bibitem{PhysRevD.24.1677}
E.~Sezgin, ``Class of ghost-free gravity Lagrangians with massive or massless
  propagating torsion,'' {  \it Phys. Rev. D}, vol.~24, pp.~1677--1680, Sep 1981.

\bibitem{Hayashi1980gravity1}
K.~Hayashi and T.~Shirafuji, ``{Gravity from Poincar{\'e} gauge theory of the
  fundamental particles. I: general formulation},'' {  \it Progress of
  Theoretical Physics}, vol.~64, pp.~866--882,  1980.

\bibitem{Hayashi1980gravity2}
K.~Hayashi and T.~Shirafuji, ``{Gravity from Poincar{\'e} gauge theory of the
  fundamental particles. II : Equations of motion for test bodies and various
  limits},'' {  \it Progress of Theoretical Physics}, vol.~64, no.~3,
  pp.~883--896, 1980.

\bibitem{Hayashi1980gravity3}
K.~Hayashi and T.~Shirafuji, ``{Gravity from Poincar{\'e} gauge theory of the
  fundamental particles. III: Weak field approximation},'' {  \it Progress of
  Theoretical Physics}, vol.~64, no.~4, pp.~1435--1452, 1980.

\bibitem{Hayashi1980gravity4}
K.~Hayashi and T.~Shirafuji, ``{Gravity from Poincar{\'e} gauge theory of the
  fundamental Particles. IV: Mass and energy of particle spectrum},'' {\it 
  Progress of Theoretical Physics}, vol.~64, pp.~2222--2241,  1980.

\bibitem{cartan1923varietes}
{\'E}.~Cartan, ``Sur les vari{\'e}t{\'e}s {\`a} connexion affine et la
  th{\'e}orie de la relativit{\'e} g{\'e}n{\'e}ralis{\'e}e (premi{\`e}re
  partie),'' {  \it Annales Sci. Ecole Norm. Sup.}, vol.~40, pp.~325--412, 1923.

\bibitem{cartan1924varietes}
{\'E}.~Cartan, ``Sur les vari{\'e}t{\'e}s {\`a} connexion affine et la
  th{\'e}orie de la relativit{\'e} g{\'e}n{\'e}ralis{\'e}e (suite),'' {  \it Annales Sci. Ecole Norm. Sup.}, vol.~41, pp.~1--25, 1924.

\bibitem{cartan1925varietes}
{\'E}.~Cartan, ``Sur les vari{\'e}t{\'e}s {\`a} connexion affine et la
  th{\'e}orie de la relativit{\'e} g{\'e}n{\'e}ralis{\'e}e (deuxi{\`e}me
  partie),'' {  \it Ann. Sci. Ecole Norm. Sup.}, vol.~42, pp.~17--88, 1925.

\bibitem{debever1979elie}
R.~Debever, J.~Leroy, and J.~Ritter, {  \it Elie Cartan--Albert Einstein: Lettres
  sur le parall{\'e}lisme absolu 1929--1932}, vol.~6 (1).
 Palais des Acad{\'e}mies {\`a} Bruxelles,  Acad{\'e}mie Royale de
  Belgique et Princeton University Press, Bruxelles (1979).

\bibitem{hehl1976general}
F.~W. Hehl, P.~Von~der Heyde, G.~D. Kerlick, and J.~M. Nester, ``General
  relativity with spin and torsion: Foundations and prospects,'' {  \it Reviews
  of Modern Physics}, vol.~48, no.~3, p.~393, 1976.

\bibitem{Palatini:1919aa}
A.~Palatini, ``Deduzione invariantiva delle equazioni gravitazionali dal
  principio di Hamilton,'' {  \it Rendiconti del Circolo Matematico di Palermo
  (1884-1940)}, vol.~43, no.~1, pp.~203--212, 1919.

\bibitem{weyl1950remark}
H.~Weyl, ``A remark on the coupling of gravitation and electron,'' {  \it Physical Review}, vol.~77, no.~5, p.~699, 1950.

\bibitem{sciama1962analogy}
D.~W. Sciama, ``On the analogy between charge and spin in general relativity,''
  {  \it Recent developments in general relativity}, Festschrift for L. Infeld, PergamonPress, Oxford; PWN, Warsaw, p.~415, 1962.
 
\bibitem{kibble1961lorentz}
T.~W. Kibble, ``Lorentz invariance and the gravitational field,'' {  \it Journal
  of mathematical physics}, vol.~2, no.~2, pp.~212--221, 1961.

\bibitem{adamowicz1980plane}
W.~Adamowicz, ``Plane waves in gauge theories of gravitation,'' {  \it General
  Relativity and Gravitation}, vol.~12, no.~9, pp.~677--691, 1980.

\bibitem{baekler1988hamiltonian}
P.~Baekler and E.~W. Mielke, ``Hamiltonian structure of Poincar{\'e} gauge
  theory and separation of non-dynamical variables in exact torsion
  solutions,'' {  \it Fortschritte der Physik/Progress of Physics}, vol.~36,
  no.~7, pp.~549--594, 1988.

\bibitem{blagojevic2017generalized}
M.~Blagojevi{\'c}, B.~Cvetkovi{\'c}, and Y.~N. Obukhov, ``Generalized plane
  waves in Poincar{\'e} gauge theory of gravity,'' {  \it Physical Review D},
  vol.~96, no.~6, p.~064031, 2017.

\bibitem{obukhov2018poincare}
Y.~N. Obukhov, ``Poincar{\'e} gauge gravity: An overview,'' {  \it International
  Journal of Geometric Methods in Modern Physics}, vol.~15, no.~supp01,
  p.~1840005, 2018.

\bibitem{Nikiforova_2018}
V.~Nikiforova and T.~Damour, ``Infrared modified gravity with propagating
  torsion: Instability of torsionfull de {S}itter-like solutions,'' {  \it Physical Review D}, vol.~97, p. 124014, Jun 2018.

\bibitem{PhysRevD.95.024013}
V.~Nikiforova, S.~Randjbar-Daemi, and V.~Rubakov, ``Self-accelerating universe
  in modified gravity with dynamical torsion,'' {  \it Phys. Rev. D}, vol.~95,
  p.~024013, Jan 2017.

\bibitem{zhytnikov1993conformally}
V.~Zhytnikov, ``Conformally invariant Lagrangians in metric-affine and
  Riemann-Cartan spaces,'' {  \it International Journal of Modern Physics A},
  vol.~8, no.~29, pp.~5141--5152, 1993.

\bibitem{minkevich2016towards}
A.~Minkevich, ``Towards the theory of regular accelerating universe in
  Riemann-Cartan space-time,'' {  \it International Journal of Modern Physics A},
  vol.~31, no.~02n03, p.~1641011, 2016.

\bibitem{Bakler:1980mu}
P.~Baekler, F.~W. Hehl, and E.~W. Mielke, ``{Vacuum Solutions With Double
  Duality Properties of a Quadratic Poincar\'e Gauge Field Theory},'' in {  \it {The Second Marcel Grossmann Meeting on the Recent Developments of General
  Relativity (In Honor of Albert Einstein)}}, Proceedings of the Conference held 5-11 July 1979, in Trieste, Italy. Edited by Remo Ruffini. Amsterdam: North-Holland Publication Co., p. 413, 1980.

\bibitem{baekler1983vacuum}
P.~Baekler, F.~Hehl, and H.~Lenzen, ``Vacuum solutions with double duality
  properties of the Poincar{\'e} gauge field theory. ii.,'' in {  \it Third
  Marcel Grossmann Meeting on General Relativity}, Edited by Hu Ning. Science Press and North Holland Publishing Co., pp.~107--128, 1983.

\bibitem{mielke1984reduction}
E.~W. Mielke, ``Reduction of the poincar{\'e} gauge field equations by means of
  duality rotations,'' {  \it Journal of mathematical physics}, vol.~25, no.~3,
  pp.~663--668, 1984.

\bibitem{wallner1991exact}
R.~Wallner, ``Exact solutions in {$U_4$} gravity. {I}. {T}he ansatz for self
  double dual curvature,'' {  \it General relativity and gravitation}, vol.~23,
  no.~6, pp.~623--639, 1991.

\bibitem{zhytnikov1996double}
V.~V. Zhytnikov, ``Double duality and hidden gauge freedom in the Poincar{\'e}
  gauge theory of gravitation,'' {  \it General Relativity and Gravitation},
  vol.~28, no.~2, pp.~137--162, 1996.

\bibitem{obukhov1989quadratic}
Y.~N. Obukhov, V.~Ponomariev, and V.~Zhytnikov, ``Quadratic Poincar{\'e} gauge
  theory of gravity: a comparison with the general relativity theory,'' {  \it General relativity and gravitation}, vol.~21, no.~11, pp.~1107--1142, 1989.

\bibitem{robinson1959solution}
I.~Robinson, ``A solution of the Maxwell-Einstein equations,'' {  \it Bull. Acad.
  Pol. Sci. Ser. Sci. Math. Astron. Phys}, vol.~7, p.~351, 1959.

\bibitem{bertotti1959uniform}
B.~Bertotti, ``Uniform electromagnetic field in the theory of general
  relativity,'' {  \it Physical Review}, vol.~116, no.~5, p.~1331, 1959.

\bibitem{cahen1968metriques}
M.~Cahen and R.~McLenaghan, ``M{\'e}triques des espaces lorentziens
  sym{\'e}triques {\`a} quatre dimensions,'' {  \it CR Acad. Sci. Paris S{\'e}r.
  AB}, vol.~266, pp.~A1125--A1128, 1968.

\bibitem{Brinkmann}
H.~W. Brinkmann, ``Einstein spaces which are mapped conformally on each
  other,'' {  \it Math. Ann.}, vol.~94, pp.~119--145, 1925.

\bibitem{breitenlohner1982positive}
P.~Breitenlohner and D.~Z. Freedman, ``Positive energy in anti-de Sitter
  backgrounds and gauged extended supergravity,'' {  \it Physics Letters B},
  vol.~115, no.~3, pp.~197--201, 1982.

\bibitem{breitenlohner1982stability}
P.~Breitenlohner and D.~Z. Freedman, ``Stability in gauged extended
  supergravity,'' {  \it Annals of Physics}, vol.~144, no.~2, pp.~249--281, 1982.

\bibitem{PhysRevD.80.104031}
V.~P. Nair, S.~Randjbar-Daemi, and V.~Rubakov, ``Massive spin-2 fields of
  geometric origin in curved spacetimes,'' {  \it Phys. Rev. D}, vol.~80,
  p.~104031, Nov 2009.

\bibitem{Nikiforova2009infrared}
V.~Nikiforova, S.~Randjbar-Daemi, and V.~Rubakov, ``Infrared modified gravity
  with dynamical torsion,'' {  \it Physical Review D}, vol.~80, no.~12,
  p.~124050, 2009.

\bibitem{Noether1918}
E.~Noether, ``Invariante Variationsprobleme,'' {  \it Nachrichten von der
  Gesellschaft der Wissenschaften zu G{\"o}ttingen, Mathematisch-Physikalische
  Klasse}, vol.~1918, pp.~235--257, 1918.

\bibitem{kosmann20n6oether}
Y.~Kosmann-Schwarzbach and L.~Meersseman, {  \it Les th{\'e}or{\`e}mes de
  {N}oether (Invariance et lois de conservation au $\text{XX}^e$ si{\`e}cle)}.
\newblock Ecole Polytechnique Universit{\'e} de Paris-Saclay, 2006.

\bibitem{lichnerowicz1955theories}
A.~Lichnerowicz, {  \it Th{\'e}ories relativistes de la gravitation et de
  l'{\'e}lectromagn{\'e}tisme: relativit{\'e} g{\'e}n{\'e}rale et th{\'e}ories
  unitaires}.
\newblock Masson, 1955.

\bibitem{nikiforova2017stability}
V.~Nikiforova, ``The stability of self-accelerating universe in modified
  gravity with dynamical torsion,'' {  \it International Journal of Modern
  Physics A}, vol.~32, no.~23n24, p.~1750137, 2017.

\bibitem{damour2019spherically}
T.~Damour and V.~Nikiforova, ``Spherically symmetric solutions in torsion
  bigravity,'' {  \it Physical Review D}, vol.~100, no.~2, p.~024065, 2019.

\bibitem{nikiforova2020black}
V.~Nikiforova and T.~Damour, ``Black holes in torsion bigravity,'' {  \it Physical Review D}, vol.~102, no.~8, p.~084027, 2020.

\bibitem{Bach:1921aa}
R.~Bach, ``Zur Weylschen Relativit{\"a}tstheorie und der Weylschen Erweiterung
  des Kr{\"u}mmungstensorbegriffs,'' {  \it Mathematische Zeitschrift}, vol.~9,
  no.~1, pp.~110--135, 1921.

\bibitem{Lanczos1938}
C.~Lanczos, ``A remarkable property of the {R}iemann-{C}hristoffel tensor in
  four dimensions,'' {  \it Annals of Mathematics, Second Series}, vol.~39,
  p.~842, Oct 1938.

\bibitem{debney1978equivalence}
G.~Debney, E.~E. Fairchild, and S.~T. Siklos, ``Equivalence of vacuum
  Yang-Mills gravitation and vacuum Einstein gravitation,'' {  \it General
  Relativity and Gravitation}, vol.~9, no.~10, pp.~879--887, 1978.

\bibitem{yang1974integral}
C.~N. Yang, ``Integral formalism for gauge fields,'' {  \it Physical Review
  Letters}, vol.~33, no.~7, p.~445, 1974.

\bibitem{ni1975yang}
W.-T. Ni, ``Yang's gravitational field equations,'' {  \it Physical Review
  Letters}, vol.~35, no.~5, p.~319, 1975.

\bibitem{pavelle1975unphysical}
R.~Pavelle, ``Unphysical solutions of Yang's gravitational-field equations,''
  {  \it Physical Review Letters}, vol.~34, no.~17, p.~1114, 1975.

\bibitem{pavelle1976unphysical}
R.~Pavelle, ``Unphysical characteristics of Yang's pure-space equations,'' {  \it Physical Review Letters}, vol.~37, no.~15, p.~961, 1976.

\bibitem{ehlers1962exact}
J.~Ehlers and W.~Kundt, ``Exact solutions of the gravitational field
  equations,'' in {  \it The Theory of Gravitation}, pp.~49--101, John Wiley \&
  Sons, Inc., 1962.

\bibitem{Penrose1965}
R.~Penrose, ``A remarkable property of plane waves in general relativity,''
  {  \it Rev. Mod. Phys.}, vol.~37, pp.~215--220, Jan 1965.

\bibitem{penrose1976any}
R.~Penrose, ``Any space-time has a plane wave as a limit,'' in {  \it Differential geometry and relativity} (M.~Cahen and M.~Flato, eds.),
  (Dordrecht, Netherlands; Boston, U.S.A.), pp.~271--275, D. Reidel Pub. Cie,,
  1976.

\bibitem{deser1975plane}
S.~Deser, ``Plane waves do not polarize the vacuum,'' {  \it Journal of Physics
  A: Mathematical and General}, vol.~8, no.~12, p.~1972, 1975.

\bibitem{minkevich1980generalised}
A.~Minkevich, ``Generalised cosmological Friedmann equations without
  gravitational singularity,'' {  \it Physics Letters A}, vol.~80, no.~4,
  pp.~232--234, 1980.

\bibitem{ramaswamy1979birkhoff}
S.~Ramaswamy and P.~B. Yasskin, ``Birkhoff theorem for an $R+R^2$ theory of
  gravity with torsion,'' {  \it Physical Review D}, vol.~19, no.~8, p.~2264,
  1979.

\bibitem{rauch1982birkhoff}
R.~Rauch, J.~C. Shaw, and H.-T. Nieh, ``Birkhoff's theorem for ghost-free,
  tachyon-free $R+ R^2+ Q^2$ theories with torsion,'' {  \it General Relativity and
  Gravitation}, vol.~14, no.~4, pp.~331--354, 1982.

\bibitem{Obukhov:2020hlp}
Y.~N. Obukhov, ``{Generalized Birkhoff theorem in the Poincar\'e gauge gravity
  theory},'' {  \it Phys. Rev. D}, vol.~102, no.~10, p.~104059, 2020.

\bibitem{dimakis1989initial}
A.~Dimakis, ``The initial value problem of the Poincar{\'e} gauge theory in
  vacuum. ii. first order formalism,'' in {  \it Annales de l'IHP Physique
  th{\'e}orique}, vol.~51, pp.~389--417, 1989.

\bibitem{hecht1996some}
R.~D. Hecht, J.~M. Nester, and V.~V. Zhytnikov, ``Some Poincar{\'e} gauge
  theory Lagrangians with well-posed initial value problems,'' {  \it Physics
  Letters A}, vol.~222, no.~1-2, pp.~37--42, 1996.

\bibitem{chen1994poincare}
C.-M. Chen, D.-C. Chern, J.~M. Nester, P.-K. Yang, and V.~V. Zhytnikov,
  ``Poincar{\'e} gauge theory Schwarzschild-de Sitter solutions with long range
  spherically symmetric torsion,'' {  \it Chinese Journal of Physics}, vol.~32,
  no.~1, pp.~29--40, 1994.

\bibitem{lenzen1985spherically}
H.-J. Lenzen, ``On spherically symmetric fields with dynamic torsion in gauge
  theories of gravitation,'' {  \it General relativity and gravitation}, vol.~17,
  no.~12, pp.~1137--1151, 1985.

\bibitem{carterCargese1978}
B.~Carter, ``Underlying mathematical structures of classical gravitation
  theory,'' {  \it Recent Developments in Gravitation: Carg{\`e}se 1978}, Edited by M. Levy and S. Deser,
  NATO Science~Series B, Physics; vol. 44, p.~41, Springer US, Plenum Press, New York 1979.


\bibitem{Zumino:1983ew}
B.~Zumino, ``{Chiral anomalies and differential geometry : lectures given at
  Les Houches, August 1983},'' in {  \it {Relativity, Groups and Topology II,
  Proceedings, 40th Summer School of Theoretical Physics - Les Houches, France,
  June 27 -- August 4, 1983}}, Edited by B. DeWitt, R. Stora, Amsterdam : North-Holland, pp.~1291--1322, 1984.

\bibitem{lovelock1971einstein}
D.~Lovelock, ``The Einstein tensor and its generalizations,'' {  \it Journal of
  Mathematical Physics}, vol.~12, no.~3, pp.~498--501, 1971.

\end{thebibliography}
\end{document}